%% file: main_NJD.tex
\documentclass[AMS,Times1COL]{WileyNJDv5} %STIX1COL,STIX2COL,STIXSMALL

\usepackage{amsmath}
\usepackage{amsfonts}
\usepackage{stmaryrd}
\usepackage{epsfig}
\usepackage{graphics} 
\usepackage{gensymb}

\usepackage{pgfplots,pgfplotstable}
\usepgfplotslibrary{groupplots}
\usetikzlibrary{shapes,arrows,snakes,calendar,matrix,backgrounds,folding,calc,positioning,spy}
\usepackage{tikz}
\usetikzlibrary{arrows.meta}
\usetikzlibrary{calc}
\tikzset{>=latex}
\usepackage[T1]{fontenc}
\usepackage{graphicx}
\usepackage{subcaption}
\usepackage{url}

\articletype{Research Article}%

\received{Date Month Year}
\revised{Date Month Year}
\accepted{Date Month Year}
\journal{Journal}
\volume{00}
\copyyear{2024}
\startpage{1}

\usepackage{todonotes}

\def\doubleunderline#1{\underline{\underline{#1}}}

\newcommand{\tr}{\operatorname{tr}}
\newcommand{\dev}{\operatorname{dev}}
\renewcommand{\div}{\operatorname{div}}

\newcommand{\xaver}[1]{{\color{black}{#1}}}

\newcommand\numberthis{\addtocounter{equation}{1}\tag{\theequation}}

\begin{document}

\title{High-order projection-based upwind method for simulation of transitional turbulent flows}

\author[1]{Philip L. Lederer}

\author[2]{Xaver Mooslechner}

\author[3]{Joachim Sch\"oberl}

\authormark{Lederer \textsc{et al.}}
\titlemark{HOPU method for simulation of transitional turbulent flows}

\address[1]{\orgdiv{Department of Applied Mathematics}, \orgname{University of Twente}, \orgaddress{\city{Enschede}, \country{The Netherlands}}}

\address[2]{\orgdiv{Institute of Analysis and Scientific Computing, Faculty of Mathematics and Geoinformation}, \orgname{TU Wien}, \orgaddress{\city{Vienna}, \country{Austria}}}

\address[3]{\orgdiv{Institute of Analysis and Scientific Computing, Faculty of Mathematics and Geoinformation}, \orgname{TU Wien}, \orgaddress{\city{Vienna}, \country{Austria}}}

\corres{Corresponding author Xaver Mooslechner, Institute of Analysis and Scientific Computing, Faculty of Mathematics and Geoinformation,
TU Wien, Vienna, Austria. \email{xaver.mooslechner@tuwien.ac.at}}

%\presentaddress{This is sample for present address text this is sample for present address text.}

%\fundingInfo{Text}
%\JELinfo{ejlje}

\abstract[Abstract]{\input{chapters/abstract}}


\jnlcitation{\cname{%
\author{Lederer P.},
\author{Mooslechner X.}, and
\author{Sch\"oberl J.}}.
\ctitle{High-order projection-based upwind method for simulation of transitional turbulent flows}
\cjournal{\it Int J Numer Meth Fluids} \cvol{year;vol(00):p--p}.}

\maketitle

\input{chapters/introduction}

\input{chapters/discretization}

\input{chapters/hopu}

\input{chapters/results}

\input{chapters/summary}

\section{Acknowledgments}
Published with the support of the Austrian Science Fund (FWF) [F65-P10 and P35931-N]. For the purpose of open access, the author has applied a CC BY public copyright license to any Author Accepted Manuscript version arising from this submission.
\\
The computational results presented have been achieved using the Vienna Scientific Cluster (VSC).

\vspace*{0.2cm}

\includegraphics[width=0.25\textwidth]{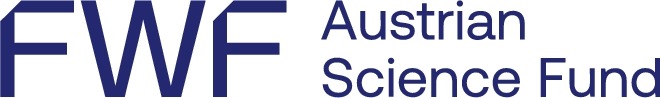}

%\renewcommand\thefootnote{}
%\footnotetext{\textbf{Abbreviations:} ANA, anti-nuclear antibodies; APC, antigen-presenting cells; IRF, interferon regulatory factor.}

%\renewcommand\thefootnote{\fnsymbol{footnote}}
%\setcounter{footnote}{1}

%\begin{thebibliography}{}
%\input{main_NJD.bib}
%\end{thebibliography}

\bibliography{literature} 

\end{document}

%% file: chapters/abstract.tex
We present a scalable, high-order implicit large-eddy simulation (ILES) approach for incompressible transitional flows. This method employs the mass-conserving mixed stress (MCS) method for discretizing the Navier-Stokes equations. The MCS method's low dissipation characteristics, combined with the introduced operator-splitting solution technique, result in a high-order solver optimized for efficient and parallel computation of under-resolved turbulent flows. We further enhance the inherent capabilities of the ILES model by incorporating high-order upwind fluxes and are examining its approximation behaviour in transitional aerodynamic flow problems. In this study, we use flows over the Eppler 387 airfoil at Reynolds numbers up to $3 \cdot 10^5$ as benchmarks for our simulations.

%% file: chapters/introduction.tex
\section{Introduction}\label{intro}

\subsection{Motivation}

Overcoming various challenges is necessary for the numerical simulation of transitional turbulent flows. As a wide range of spatial and temporal scales occur in turbulent flows, the numerical approximation of such a flow state is computationally demanding. Moreover, the prediction of intricate physical phenomenons that emerge when a flow changes from a laminar to a turbulent regime often poses difficulties.

Despite significant advancements in computational power over recent decades, complete capture of information about turbulent flows via direct numerical simulation (DNS) is not feasible for most practical applications. The large eddy simulation (LES) \cite{Sagaut} is the most commonly used alternative to this costly technique. The core concept behind LES is that large-scale energy-containing eddies dominate turbulent transport, so explicitly resolving those scales in numerical simulations provides a realistic depiction of the flow. The effects of the unresolved fine scale part of the flow are typically modeled using explicit subgrid-scale (SGS) models. LES has become a well-established discipline in computational engineering, supported by the development of numerous SGS models developed over the years. A more sophisticated approach to LES is the variational multiscale simulation (VMS) \cite{Hughes2000}, which allows for different treatment of turbulent flow scales, with the SGS model influencing only a subset of the resolved scales.

An alternative approach to incorporating dissipation effects, stemming from the absence of resolved scales, is through the use of numerical dissipation inherent in the discretization scheme. This concept, known as implicit large eddy simulation (ILES), was initially proposed by Boris in 1990 \cite{Boris1990}. While it has been employed across various numerical methods, ILES has garnered significant attention in the past decade, particularly in conjunction with discontinuous Galerkin (DG) methods \cite{Lehrenfeld2019, Peraire2017, Uranga2011, Moura2017, Bolemann2015, Gassner2013, Plata2018, deWiart2015}.

Due to their robust mathematical foundations and flexible nature, DG methods have demonstrated exceptional effectiveness across various fields of numerical computation. They offer exact conservation properties and provide stable discretizations of convective operators. Importantly, the DG framework supports the implementation of solution methods on complex and curved geometries, making it highly valuable for engineering applications. Moreover, the selective application of linear dissipation and dispersion to high-frequency components of the solution makes high-order DG methods particularly suitable for multiscale simulations. This motivates the consideration of high-order methods for addressing transitional flow problems.

The exponential amplification of small perturbations within the laminar boundary layer plays a critical role in the transition to turbulence. These perturbations typically have amplitudes several orders of magnitude smaller than the free-stream velocity, eventually leading to non-linear breakdown into a fully turbulent flow regime. It is crucial to avoid excessive dissipation of these small perturbations, as accurate prediction of the transition relies on capturing their evolution faithfully. Studies by \cite{Park2017, Fernandez2017} demonstrate ILES of transitional aerodynamic flows using high-order DG methods at Reynolds numbers of approximately $10^5$ or higher. Despite relatively coarse spatial resolutions of the discretized domain, their simulations showed good agreement with experimental data from NASA \cite{Nasa1988}.

In this work, we want to extend the results of \cite{HOPU2023} by examining the behaviour of the high-order projection-based upwind method (HOPU) for ILES in transitional flow. In this approach, numerical dissipation from the Riemann solver of the convection operator is applied only to the high-order part of the velocity solution, similar to VMS for explicit models. This novel method has shown promising results in wall-bounded flows. In \cite{Burman2007}, a projection-based upwind method was first introduced as a DG stabilization method for hyperbolic systems and later established for diffusion equations in mixed form in \cite{Burman2009}. Recently, jump penalty stabilization methods for under-resolved turbulence in DG have sparked interest \cite{Ferrer2023,Moura2022,Tong2022}. This work aims to provide a deeper understanding of these mechanisms while contributing to ongoing research into ILES for turbulent flows.

Second, given that higher Reynolds number flows require significantly more computational effort, we introduce a fully parallel DG solver to address this issue. We first consider the hybrid version of the exactly mass-conserving mixed stress method (MCS) introduced in \cite{lederer} and already used in \cite{HOPU2023} for solving the Navier-Stokes equations. Since the scalability of solving the given mixed problem is limited, we employ a decoupling of velocity and pressure using the projection method of Chorin. This approach leverages the parallel solvability of the pressure-Poisson equation and velocity prediction step through operator splitting, enabling a highly efficient and scalable algorithm. The numerical results have been obtained by using the open-source finite element framework NGSolve/Netgen \cite{ngsolve} and the scalable parallel solver package PETSc \cite{PETSC}.

\subsection{Overview}

The paper is organized as follows. In Section \ref{discretization}, we describe the numerical discretization of the Navier–Stokes equations and discuss the solution methods for the non-linear system of equations arising from the discretization. In Section \ref{methods}, we explain the HOPU approach and discuss its possible effects on transitional flows. In Sections \ref{results}, the proposed approaches are applied to transient ﬂows over the Eppler 387 wing. There, a comparison of standard upwind method to HOPU is given. Section \ref{conclusion} summarizes the observations made in this work and draws conclusions.

\subsection{Mathematical Model}

Let $\Omega \subset \mathbb{R}^{3}$ be a connected and bounded
Lipschitz domain and $[0,T_{e}]$ a given time interval with end time
$T_e \in\mathbb{R}^+$. We consider the velocity field $\underline{u}$
and the pressure $p$ as the solution of the incompressible
Navier-Stokes equations
	\begin{subequations} \label{NSE}
		\begin{alignat}{2}
	\nabla\,\cdot\,\underline{u} &=0  && \qquad\textnormal{in}\:\Omega\times[0,T_{e}], \label{NSE_C} \\
	\frac{\partial\underline{u}}{\partial t} + (\underline{u}\,\cdot\,\nabla)\underline{u}
	 - 2\nu\nabla\,\cdot\,\epsilon(\underline{u}) + \nabla p &=0 && \qquad \textnormal{in}\:\Omega\times[0,T_{e}], \label{NSE_M} \\
	\underline{u} &=\underline{u}_{0} && \qquad \textnormal{in}\:\Omega, t=0, \label{NSE_I}\\
	\frac{\partial \underline{u}}{\partial \underline{n}} &=0 &&\qquad \textnormal{in}\:\partial\Omega_{N}\times[0,T_{e}], \label{NSE_N}\\
	\underline{u} &=0 &&\qquad \textnormal{in}\:\partial\Omega_{W}\times[0,T_{e}], \label{NSE_W}\\
	\underline{u} &=\underline{u}_I &&\qquad \textnormal{in}\:\partial\Omega_{I}\times[0,T_{e}], \label{NSE_In}
\end{alignat}
\end{subequations}
where $\underline{u}_{0}$ is a compatible initial condition. The
boundary is split into four parts such that $\partial \Omega = \partial
\Omega_W \cup \partial\Omega_{I} \cup \partial \Omega_N \cup \partial \Omega_P$. We prescribe Dirichlet boundary conditions with the value $\underline{u}=0$ on $\partial\Omega_{W}$ (no-slip) or $\underline{u}=\underline{u}_I$ on $\partial\Omega_{I}$ (inlet), and on
$\partial\Omega_{P}$ we consider periodic boundary conditions. On the subpart $\partial \Omega_N$ we denote zero Neumann boundary conditions (outlet) given the normal vector $\underline{n}$ defined on $\partial \Omega$. In
addition, we denote by $\nu\in \mathbb{R}^+\setminus \{ 0 \}$ the constant
kinematic viscosity. In \eqref{NSE_M}, the term $\epsilon
(\underline{u}) = \frac{1}{2}(\nabla\underline{u} +
\nabla\underline{u}^{t})$ is the (symmetric) strain rate tensor and
$\nabla\,\cdot\,\epsilon(\underline{u})$ indicates the divergence
applied to each row of $\epsilon(\underline{u})$.

%% file: chapters/discretization.tex
\section{Solver for turbulent incompressible flows}\label{discretization}

\subsection{Notation}

Let $\mathcal{T}_h$ be a subdivision of the domain $\Omega$ into non-overlapping hexahedral elements. The skeleton $\mathcal{F}_h$ represents the set of all element facets, which is divided into three sets $\mathcal{F}_h =
\mathcal{F}^I_h \cup \mathcal{F}^W_h \cup \mathcal{F}^N_h$. Here $\mathcal{F}^I_h$ includes
all interior, periodic and non-zero Dirichlet boundary facets, $\mathcal{F}^W_h$
consists of facets on the homogeneous Dirichlet boundary and $\mathcal{F}^N_h$ comprises facets on the Neumann boundary.
Let $\underline{\psi}$ be a smooth, element-wise vector-valued function. For any elements $T, T' \in
\mathcal{T}_h$ sharing a common facet $E \in \mathcal{F}_h$, define
$\underline{\psi}^T = \underline{\psi}|_T$ and $\underline{\psi}^{T'}
= \underline{\psi}|_{T'}$. The jump
$\llbracket\cdot\rrbracket$ and average $\langle\cdot\rangle$
operators are defined as
\begin{align*}
% & \llbracket\underline{\psi}\rrbracket = \underline{\psi}|_E - \underline{\psi}^-|_E, & \langle\underline{\psi}\rangle = \frac{1}{2}(\underline{\psi}|_E + \underline{\psi}^-|_E). &
\llbracket\underline{\psi}\rrbracket = \underline{\psi}^T|_E - \underline{\psi}^{T'}|_E, \quad \textrm{and} \quad \langle\underline{\psi}\rangle = \frac{1}{2}(\underline{\psi}^T|_E + \underline{\psi}^{T'}|_E).
\end{align*}
On the element boundary $\partial T$, the tangential trace operator is
given by
\begin{equation*}
% \underline{\psi}_t = (\underline{\psi}|_E-(\underline{\psi}|_E\cdot\underline{n})\underline{n}).
\underline{\psi}_t = \underline{\psi}-(\underline{\psi}\cdot\underline{n})\underline{n},
\end{equation*}
where $\underline n$ is the outward normal vector.
Additionally, the normal-normal and normal-tangential
trace operators for smooth matrix-valued functions
$\doubleunderline{\psi}$ are introduced as
\begin{align*}
 \doubleunderline{\psi}_{nn} = (\doubleunderline{\psi} n)\cdot n, \quad \textrm{and} \quad \doubleunderline{\psi}_{nt} = \doubleunderline{\psi}n-(\doubleunderline{\psi}_{nn})n.
\end{align*}
The $\mathrm{skew}(\cdot)$ operator and the deviatoric part of a matrix are given as
\begin{equation*}
\mathrm{skew}(\doubleunderline{\psi})=\frac{1}{2} ( \doubleunderline{\psi} - \doubleunderline{\psi}^t ), \quad \textrm{and} \quad
\dev({\doubleunderline{\psi}}) =
\doubleunderline{\psi} - \frac{1}{3} \tr(\doubleunderline{\psi}) I,
\end{equation*}
where $\tr(\cdot)$ denotes the matrix trace and $I$ is the identity matrix.

In this work, $H^1(\Omega, \mathbb{R})$ represents the standard first-order Sobolev
space, whereas $H^1(\Omega, \mathbb{R}^3)$ denotes its
vector-valued version.
The Sobolev space $H(\mathrm{div},\xaver{\Omega})$ is defined as
\begin{equation*}
H(\mathrm{div},\Omega)=\{\underline{v}\in L^2(\Omega,\mathbb{R}^3):\nabla\,\cdot\,\underline{v}\in L^2(\Omega,\mathbb{R})\}.
\end{equation*}
For all the above Sobolev spaces, a zero subscript indicates
that the corresponding continuous trace vanishes on $\partial
\Omega_D = \partial
\Omega_W \cup \partial \Omega_I$. For example, $H_0(\div, \Omega)$ denotes all functions $\underline{v} \in
H(\div, \Omega)$ such that $\underline{v}_n = 0$ on $\partial \Omega_D$. For
conforming (normal continuous) approximations of functions in
$H(\mathrm{div},\Omega)$, we consider the Raviart-Thomas finite element
space $RT^k(\mathcal{T}_h)$, as described in \cite{raviart, boffi2013mixed}, where
$k$ is a given order. Additionally, $\mathbb{P}^{k}(\mathcal{T}_h, \mathbb{R})$
denotes the space of all
element-wise polynomials defined on $\mathcal{T}_h$ with total order
less than or equal $k$.

\subsection{The discrete model - the MCS method}\label{mcs}

In this work, we consider the so called mass conserving
mixed stress (MCS) method, which is detailed in \cite{HOPU2023} and originates from \cite{ledererthesis,
lederer, lederer2}. Since we discuss only a slightly different variant (incorporating non-homogeneous Dirichlet
boundary conditions) compared to the one in \cite{HOPU2023}, we will focus on the most
critical aspects of the method. The core idea of MCS is to reformulate the incompressible
Navier-Stokes equations into a first-order formulation that
includes the stress.

Introducing the auxillary variables $\doubleunderline{\gamma} =
{\omega}(\underline u) = \mathrm{skew}(\nabla \underline{u})$, $\doubleunderline{\sigma} =
2\nu\epsilon(\underline{u})$ and the relation $\nabla \underline{u} = 
\epsilon(\underline{u}) + {\omega}(\underline u)$, we can rewrite the original system
\eqref{NSE} as
\begin{subequations} \label{MCS_cont}
\begin{alignat}{2}
	-\frac{1}{2\nu}{\doubleunderline{\sigma}} + \nabla\underline{u} - \doubleunderline{\gamma} &= \doubleunderline{0} && \qquad \textnormal{in}\:\Omega\times[0,T_e],  \label{MCS_cont_S}\\
	\frac{\partial\underline{u}}{\partial t} - \nabla\cdot\doubleunderline{\sigma} + (\underline{u}\,\cdot\,\nabla)\underline{u} + \nabla p &= \underline{0} && \qquad \textnormal{in}\:\Omega\times[0,T_e], \label{MCS_cont_M} \\
\nabla\,\cdot\,\underline{u}&=0 && \qquad\textnormal{in}\:\Omega\times[0,T_e], \\
\mathrm{skew}(\doubleunderline{\sigma})&=\doubleunderline{0} && \qquad \textnormal{in}\:\Omega\times[0,T_e], \\
	\underline{u}&=\underline{u}_{0} && \qquad \textnormal{in}\:\Omega, t=0,  \\
	\frac{\partial \underline{u}}{\partial \underline{n}} &=\underline{0} &&\qquad \textnormal{in}\:\partial\Omega_{N}\times[0,T_{e}]\\
	\underline{u}&=0 & & \qquad \textnormal{in}\:\partial\Omega_{W}\times[0,T_e],\\
	\underline{u}&=\underline{u}_I & & \qquad \textnormal{in}\:\partial\Omega_{I}\times[0,T_e].
\end{alignat}
\end{subequations}
The additional constraint $\mathrm{skew}(\doubleunderline{\sigma})=0$
ensures symmetry of $\doubleunderline{\sigma}$ and is used to
enforce weak symmetry of the approximate stress, as common for mixed finite element methods (see \cite{FdV2, PEERS, family, ls2}).
By selecting the finite element spaces for the
velocity and pressure by
\begin{align} \label{disc_velspace}
 V_h= RT^k(\mathcal{T}_h) \cap H_{0}(\mathrm{div}, \Omega), \quad \textrm{and} \quad  Q_h= \mathbb{P}^{k}(\mathcal{T}_h) \subset L^2(\Omega),
\end{align}
we have the compatibility condition
\begin{align} \label{divVeqQ}
	\nabla\,\cdot\,V_h \subseteq Q_h.
\end{align}
The choice of the (discrete) velocity and
pressure space results in exactly divergence-free velocity
approximations (see \cite{boffi2013mixed}). However, due to the missing tangential
continuity, $V_h$ is not a subset of $H^1(\Omega, \mathbb{R}^3)$ and thus, the gradient operator in \eqref{MCS_cont_S} is not
well-defined for discrete velocity functions ${\underline{u}}_h \in
V_h$.

This motivates our redefinition of the above equations
as a first-order system. By multiplying equation \eqref{MCS_cont_S} with a
stress test function $\doubleunderline{\tau}$ and integrating
over the domain $\Omega$, integration by parts for the second term yields
\begin{align} \label{eq:contfirstline}
	-\frac{1}{\nu} (\doubleunderline{\sigma},\doubleunderline{\tau})
	- (\nabla \cdot \doubleunderline{\tau}, \underline{u})
	- (\doubleunderline{\gamma}, \doubleunderline{\tau}) = 0.
\end{align}
The primary finding here is that the second term is now well-defined
for functions in $H(\div)$. The regularity has shifted from the
velocity $\underline{u}$ to the stress test function
$\doubleunderline{\tau}$. A detailed analysis to further motivate this is 
necessary but omitted here for simplicity.
For more details, we refer to \cite{ledererthesis, lederer}.

An approximation space for (discrete) stresses/strains resulting in a
stable method (in respect with the spaces $V_h$ and $Q_h$) is given
by normal-tangential continuous polynomials of a certain
order
\begin{equation*}
	\{ \doubleunderline{\tau}_h \in \mathbb{P}^{k+1}(\mathcal{T}_h,\mathbb{R}^{3 \times 3}): \, \tr(\doubleunderline{\tau}_h) = 0,   (\doubleunderline{\tau}_h)_{nt} \in \mathbb{P}^k(E, \mathbb{R}^3),  \llbracket  (\doubleunderline{\tau}_h)_{nt} \rrbracket = 0 \,  \forall E\in\mathcal{F}_h^I\}.
\end{equation*}
This space consists of element-wise, matrix-trace-free, matrix-valued polynomials of total order $k+1$,
whose normal-tangential trace on facets is a vector-valued polynomial of order $k$ and whose
normal-tangential jump vanishes.

Despite incorporating normal-tangential
continuity into the space as described above, we use an additional
hybrid variable that serves
as a Lagrange multiplier (see \cite{refId0}). This has the advantage of decoupling the
local element-wise degrees of freedom of the stress approximation, allowing them to be
locally eliminated (see Section 4.2 in \cite{lederer3}).

Overall, this leads to the definition of
\begin{align*}
	\Sigma_h &=\{ \doubleunderline{\tau}_h \in \mathbb{P}^{k+1}(\mathcal{T}_h,\mathbb{R}^{3 \times 3}): \tr(\doubleunderline{\tau}_h) = 0,   (\doubleunderline{\tau}_h)_{nt} \in \mathbb{P}^k(E, \mathbb{R}^3)\; \forall E\in\mathcal{F}_h^I\},\\
	\hat{V}_h&=\{\hat{\underline{v}}_h\in \mathbb{P}^{k}(\mathcal{F}_h, \mathbb{R}^3):\hat{\underline{v}}_h\,\cdot\,\underline{n}=0 \; \forall E\in\mathcal{F}_h, \; \hat{\underline{v}}_h = 0 \; \forall E\in\mathcal{F}^D_h\}.
\end{align*}
Note that with this choice, the normal-tangential jump of functions from the "broken" stress space $\Sigma_h$
are elements of $\hat{V}_h$, i.e.,
\begin{align} \label{jumpinVhat}
	\llbracket  (\doubleunderline{\tau}_h)_{nt} \rrbracket \in \hat V_h \quad \forall \tau_h \in \Sigma_h, \forall E \in \mathcal{F}_h.
\end{align}

It remains to define a space for the approximation of
$\doubleunderline{\gamma}$, which is given by
\begin{equation*}
W_h=\mathbb{P}^{k}_{skew}(\mathcal{T}_h,\mathbb{R}^{3 \times 3}) =
\{ \doubleunderline{\eta} \in
\mathbb{P}^{k}(\mathcal{T}_h,\mathbb{R}^{3 \times 3}) :
(\doubleunderline{\eta} + \doubleunderline{\eta}^t)= 0 \: \forall
T\in\mathcal{T}_h \}.
\end{equation*}

The (semi-)discrete hybrid MCS method with weakly imposed symmetry then reads as:
Find
$(\doubleunderline{\sigma}_h,\underline{u}_h,\underline{\hat{u}}_h,\doubleunderline{\gamma}_h,p_h)
\in \Sigma_h \times V_h \times \hat{V}_h \times W_h \times Q_h$ such
that
\begin{subequations} \label{MCS}
\begin{alignat*}{2}
a_h (\doubleunderline{\sigma}_h,\doubleunderline{\tau}_h) + b_{2h}(\doubleunderline{\tau}_h,(\underline{u}_h,\underline{\hat{u}}_h,\doubleunderline{\gamma}_h))&=0 & & \qquad\forall \doubleunderline{\tau}_h \in \Sigma_h, \numberthis \label{MCS_sigma} \\
(\frac{\partial \underline{u}_h}{\partial t},\underline{v}_h) + b_{2h}(\doubleunderline{\sigma}_h,(\underline{v}_h,\underline{\hat{v}}_h,\doubleunderline{\eta}_h)) \numberthis \label{MCS_u} \hspace{1cm}& & & \\
+ b_{1h}(\underline{v}_h,p_h)+c_h(\underline{u}_h,\underline{u}_h,\underline{v}_h)&=0 & &\qquad\forall (\underline{v}_h,\underline{\hat{v}}_h, \doubleunderline{\eta}_h) \in V_{h} \times \hat{V}_{h} \times W_h, \\
b_{1h}(\underline{u}_h,q_h)&=0 \numberthis \label{MCS_p}&& \qquad \forall q_h\in Q_h.
\end{alignat*}
\end{subequations}
The bilinear forms are defined as follows. The symmetric form, which
includes the stress variables and the incompressibility constraint,
is given by
\begin{align*}
a_h (\doubleunderline{\sigma}_h,\doubleunderline{\tau}_h) &= \sum_{T\in\mathcal{T}_h} \int_T -\frac{1}{2\nu}\:\doubleunderline{\sigma}_h\,:\,\doubleunderline{\tau}_h\:d\underline{x},
\\
% \end{equation}
b_{1h}(\underline{u}_h,q_h) &=-\sum_{T\in\mathcal{T}_h}\int_T(\nabla\,\cdot\,\underline{u}_h)q_h\:d\underline{x} = -(\nabla\,\cdot\,\underline{u}_h,q_h),
\end{align*}
respectively, where the last equation follows since $\underline{u}_h$
is normal continuous. Note that due to \eqref{divVeqQ}, equation
\eqref{MCS_p} enforces an exactly divergence-free property of the
discrete velocity solution $\underline u_h$.
The other constraint is given by
\begin{align*}
b_{2h}(\doubleunderline{\sigma}_h,(\underline{v}_h,\underline{\hat{v}}_h,\doubleunderline{\eta}_h))
= \numberthis \label{b2h} & \sum_{T\in\mathcal{T}_h}
\bigg( -\int_T (\nabla\,\cdot\,\doubleunderline{\sigma}_h)\,\cdot\,\underline{v}_h\:d\underline{x}
+ \int_{\partial T}  (\doubleunderline{\sigma}_h)_{nn} (\underline{v}_h\,\cdot\,\underline{n})\:d\underline{s}
\\ & - \int_T \doubleunderline{\sigma}_h\, : \, \doubleunderline{\eta}_h\:d\underline{x}
+ \int_{\partial T}  (\doubleunderline{\sigma}_h)_{nt}\,\cdot\,\underline{\hat{v}}_h\:d\underline{s} \bigg).
\end{align*}
To motivate this, consider the first line \eqref{MCS_sigma} for
an arbitrary test function $\doubleunderline{\tau}_h$,
\begin{align*}
	a_h (\doubleunderline{\sigma}_h,\doubleunderline{\tau}_h) + b_{2h}(\doubleunderline{\tau}_h,(\underline{u}_h,\underline{\hat{u}}_h,\doubleunderline{\gamma}_h))=0,
\end{align*}
and compare it with the continuous version \eqref{eq:contfirstline}.
While $a_h (\doubleunderline{\sigma}_h,\doubleunderline{\tau}_h)$ is
the same $1/(2\nu)$-scaled $L^2$-inner product as the first term in
\eqref{eq:contfirstline}, the sum
\begin{align*}
	\sum_{T\in\mathcal{T}_h}
 -\int_T (\nabla\,\cdot\,\doubleunderline{\tau}_h)\,\cdot\,\underline{u}_h\:d\underline{x}
+ \int_{\partial T}  (\doubleunderline{\tau}_h)_{nn} (\underline{u}_h\,\cdot\,\underline{n})\:d\underline{s},
\end{align*}
corresponds to a discrete (distributional) version of $-(\nabla \cdot
\doubleunderline{\tau}, u)$. The third integral in the term
$b_{2h}(\doubleunderline{\tau}_h,(\underline{u}_h,\underline{\hat{u}}_h,\doubleunderline{\gamma}_h))$
corresponds to the third term in~\eqref{eq:contfirstline} and is
needed to incorporate (weak) symmetry. The remaining integral includes
the Lagrange multiplier $\hat{\underline u}_h$, which is the symmetric
term given in \eqref{MCS_u}. This term enforces
normal-tangential continuity of $\doubleunderline{\sigma}_h$, since
(from~\eqref{MCS_u}) we have
\begin{align*}
	\sum_{T\in\mathcal{T}_h} \int_{\partial T}  (\doubleunderline{\sigma}_h)_{nt}\,\cdot\,\underline{\hat{v}}_h\:d\underline{s}
	= \sum_{F \in \mathcal{F}_h} \int_F \llbracket  (\doubleunderline{\sigma}_h)_{nt} \rrbracket \,\cdot\,\underline{\hat{v}}_h\:d\underline{s} = 0 \quad \forall \underline{\hat{v}}_h \in \hat V_h,
\end{align*}
and thus by \eqref{jumpinVhat} we have $\llbracket  (\doubleunderline{\sigma}_h)_{nt} \rrbracket = 0$.

The non-linear convective term is the same as in \cite{lehrenfeld} and is
given by
\begin{align}
c_h(\underline{u}_h,\underline{u}_h,\underline{v}_h)= & \sum_{T\in\mathcal{T}_h}\int_T(\nabla\underline{u}_h\cdot\underline{u}_h)\cdot\underline{v}_h\:d\underline{x} 
 +\sum_{E\in\mathcal{F}^I_h} \bigg(- \int_{E}(\underline{u}_h\,\cdot\,\underline{n})\llbracket\underline{u}_h\rrbracket\,\cdot\,\langle\underline{v}_h\rangle\:d\underline{s} %\nonumber \\
+\frac{1}{2}|\underline{u}_h\,\cdot\,\underline{n}|\llbracket\underline{u}_h\rrbracket\,\cdot\,\llbracket\underline{v}_h\rrbracket\:d\underline{s}\bigg). \label{upwind} %\nonumber
\end{align}
On $E\in\mathcal{F}^I_h \cap \partial \Omega_I$, the relation $\underline{u}_h^T|_E-\underline{u}_I|_E$ is used for the jump and analogously
for the average operator.
The last integral results in a standard (DG) upwind stabilization, as discussed in \cite{MR2165335}.
If this term is omitted, the method lacks convection stabilization and relies instead on a central flux approach.

In \eqref{MCS}, the Dirichlet no-slip conditions in \eqref{MCS_cont} have been decomposed
into normal and tangential components. While $\underline u_h \cdot
\underline n = 0$ is enforced within the space $V_h$ (refer to
\eqref{disc_velspace}), the requirement for the tangential trace to vanish is imposed through the facet variables in $\hat{V}_h$.
A comprehensive discussion on these
boundary conditions, including non-zero Neumann and mixed conditions, can be found
in Section 4 of \cite{ledererthesis}.

Lastly, it is important to emphasize that numerical quadrature is applied to all bilinear forms
to ensure exact integration. This practice is crucial for maintaining the stability of the method,
especially in turbulent flow regimes.

\subsection{Operator splitting method}\label{splitting}
As in \cite{HOPU2023}, one can use the Runge-Kutta scheme of the implicit-explicit method (IMEX) \cite{IMEX} to derive a fully discrete method.
There, the convection term \eqref{upwind} is always treated explicitly while the Stokes problem, composed of the bilinear forms $a_h(\cdot,\cdot)$, $b_{2h}(\cdot,\cdot)$ and $b_{1h}(\cdot, \cdot)$, is treated implicitly to preserve the precise divergence-free property of the discrete velocity. Therefore, the need for a non-linear solver is negated.

In this work we present a method to solve the system from \eqref{MCS} based on projections. Originally presented by Chorin \cite{Chorin1968}, this technique allows for decoupling of pressure and velocity, thereby reduces the complexity of numerically solving the saddle-point structured problem.
In the projection method, the incompressibility constraint is initially ignored and the momentum equation (without the pressure term) is used to compute an intermediate velocity field $\underline{u}_h^*$. Then, to obtain the new updated solution $\underline{u}_h^{n+1}$, the velocity is 
projected back to the incompressible velocity space.

The splitting scheme for MCS, here for simplicity shown for the first order in time, can be summarized in two steps. Beginning with the first step, find
$(\doubleunderline{\sigma}_h,\underline{u}_h^*,\underline{\hat{u}}_h,\doubleunderline{\gamma}_h)
\in \Sigma_h \times V_h \times \hat{V}_h \times W_h$ such that
\begin{subequations} \label{MCS_step1}
\begin{align*}
a_h (\doubleunderline{\sigma}_h,\doubleunderline{\tau}_h) + b_{2h}(\doubleunderline{\tau}_h,(\underline{u}_h^*,\underline{\hat{u}}_h,\doubleunderline{\gamma}_h))&=0 & & \forall \doubleunderline{\tau}_h \in \Sigma_h, \numberthis  \\
(\frac{\underline{u}_h^*-\underline{u}_h^n}{\Delta t},\underline{v}_h) + b_{2h}(\doubleunderline{\sigma}_h,(\underline{v}_h,\underline{\hat{v}}_h,\doubleunderline{\eta}_h)) \numberthis  & & & \\
 +c_h(\underline{u}_h^n,\underline{u}_h^n,\underline{v}_h) &=0 & &\forall (\underline{v}_h,\underline{\hat{v}}_h, \doubleunderline{\eta}_h) \in V_{h} \times \hat{V}_{h} \times W_h,
\end{align*}
\end{subequations}
given $\underline{u}_h^n$ and time step $\Delta t$.
Similarly to the IMEX approach, the stiff part, such as diffusive terms, is treated implicitly while convection is handled explicitly. The incompressibility term (bilinear form $b_{1h}(\cdot,
\cdot)$) is left out, so the intermediate velocity $\underline{u}_h^*$ is not divergence-free.
After obtaining the intermediate solution, we project $\underline{u}_h^*$ back into the space of exactly divergence-free velocities, resulting in $\underline{u}_h^{n+1}$. For the second step, find $(\underline{\tilde{u}}_h,p_h,\hat{p}_h) \in V_h^{disc.} \times Q_h \times \hat{Q}_h$ such that
\begin{subequations} \label{MCS_step2}
\begin{align*}
-(\underline{\tilde{u}}_h,\underline{\tilde{v}}_h)+b_{1h}(\underline{\tilde{v}}_h,p_h)+b_{3h}(\underline{\tilde{v}}_h,\hat{p}_h)&=0 & & \forall \underline{\tilde{v}}_h \in V_h^{disc.}, \numberthis  \\
b_{1h}(\underline{\tilde{u}}_h,q_h)+b_{3h}(\underline{\tilde{u}}_h,\hat{q}_h) &=b_{1h}(\underline{u}_h^*,q_h) & &\forall (q_h, \hat{q}_h) \in Q_{h} \times \hat{Q}_h, \numberthis \label{MCS_step2_2}
\end{align*}
\end{subequations}
whereas the finite element spaces are defined as
\begin{equation*}
V_h^{disc.} = RT^{k,disc.}(\mathcal{T}_h)
= \{v_h \in L^2(\Omega, \mathbb{R}^d): v_h|_T \in RT^k(T) \, \forall T \in \mathcal{T}_h \}
,
\end{equation*}
\xaver{defined as the space of normal discontinuous velocities} (here $RT^k(T)$ is the local Raviart-Thomas space of order $k$ on an element $T \in \mathcal{T}_h$) and
\begin{equation*}
\hat{Q}_h = \{\hat{q}_h\in \mathbb{P}^{k}(\mathcal{F}_h,\mathbb{R}):\hat{q}_h = 0 \; \forall E\in\mathcal{F}^N_h\}.
\end{equation*}
Since the vector-valued space $V_h^{disc.}$ is no longer normal continuous, continuity is weakly enforced by
\begin{equation*}
b_{3h}(\underline{\tilde{u}}_h,\hat{q}_h) = \sum_{T\in\mathcal{T}_h} \int_{\partial T}  (\underline{\tilde{u}}_h\,\cdot\,\underline{n})\hat{q}_h\:d\underline{s}.
\end{equation*}
Although enforced weakly, note that this gives $\llbracket \underline{\tilde{u}}_h\,\cdot\,\underline{n} \rrbracket = 0$ in a point-wise sense since (by the choice of spaces) choosing 
$q_h = 0$ and $\hat{q}_h = \llbracket \underline{\tilde{u}}_h\,\cdot\,\underline{n} \rrbracket \in 
\hat{Q}_h$, we obtain by \ref{MCS_step2_2} 
\begin{equation*}
	\sum_{F\in\mathcal{F}^I_h}\int_F  \llbracket \underline{\tilde{u}}_h\,\cdot\,\underline{n} \rrbracket ^2 \: d\underline{s} = 0,
	\end{equation*}
and thus $\underline{\tilde{u}}_h \in V_h$. We then define the update
as
\begin{equation}
\underline{u}_h^{n+1} = \underline{u}_h^{n} + \Delta t(\underline{u}_h^* -
\underline{\tilde{u}}_h).
\label{UpdateStep}
\end{equation}
%To further elaborate that scheme \eqref{MCS_step2} is indeed a projection we recall
%the $L^2$-orthogonal Helmholtz decomposition. This implies
%that every function $\underline{\psi}^* \in L^2(\Omega,\mathbb{R}^3)$ can be uniquely decomposed into
%\begin{equation*}
%\underline{\psi}^* = \underline{\psi} + \nabla \tilde{\psi} ,
%\end{equation*}
%with $\nabla \cdot \underline{\psi} = 0$ and $\underline{\psi} \cdot \underline{n} |_{\partial
% \Omega} = 0$. Setting $\underline{\psi}^* := \underline{u}_h^*$ and $\nabla \tilde{\psi}
%:= \nabla p_h = \underline{\tilde{u}}_h$ gives exactly the last term in equation 
%\eqref{UpdateStep}.
The updated velocity is indeed pointwise divergence-free since for $\underline{u}_h^* \in V_h$ and, following above observations, $\underline{\tilde{u}}_h \in V_h$ we can test
equation \ref{MCS_step2_2} for any $q_h \in Q_h$ and $\hat{q}_h = 0$ to get 
\begin{equation*}
\sum_{T\in\mathcal{T}_h}\int_T\big(\nabla\,\cdot\,(\underline{u}_h^*-\underline{\tilde{u}}_h)\big)q_h\:d\underline{x} = 
(\nabla\,\cdot\,(\underline{u}_h^*-\underline{\tilde{u}}_h), q_h) = 0,
\end{equation*}
thus again by \eqref{divVeqQ} we have $\nabla\,\cdot\,(\underline{u}_h^*-\underline{\tilde{u}}_h) = 0$.

Denoting $\phi_{\doubleunderline{\sigma}}$, $\phi_{\doubleunderline{\gamma}}$, $\phi_{\underline{u}}$ and $\phi_{\underline{\hat{u}}}$ as the basis functions of $\Sigma_h$ , $W_h$ , $V_h$ and $\hat{V}_h$ respectively and, complying with the notation for the Galerkin isomorphism, $\textbf{\underline{u}}$ for the coefficients of $\underline{u}_h$ with respect to the basis given by $\phi_{\underline{u}}$ , etc., the system of equations
\eqref{MCS_step1} in matrix form is
\begin{equation}
\begin{pmatrix}
M_{\doubleunderline{\sigma}\doubleunderline{\sigma}} & B_{\doubleunderline{\gamma}\doubleunderline{\sigma}}^t & B_{\underline{u}\doubleunderline{\sigma}}^t & B_{\underline{\hat{u}}\doubleunderline{\sigma}}^t \\
B_{\doubleunderline{\gamma}\doubleunderline{\sigma}} & 0 & 0 & 0 \\
B_{\underline{u}\doubleunderline{\sigma}} & 0 & M_{\underline{u}\underline{u}} & 0 \\
B_{\underline{\hat{u}}\doubleunderline{\sigma}} & 0 & 0 & 0
\end{pmatrix}
\begin{pmatrix}
\mathbf{\doubleunderline{\sigma}}  \\
\mathbf{\doubleunderline{\gamma}}  \\
\mathbf{\underline{u}^*}  \\
{\underline{\hat{\mathbf{u}}}}
\end{pmatrix}
=
\begin{pmatrix}
0  \\
0  \\
M_{\underline{u}\underline{u}}\mathbf{\underline{u}^{n}}-C_{\underline{u}\underline{u}}\mathbf{\underline{u}^{n}}  \\
0
\end{pmatrix},
\end{equation}
where
\begin{align*}
M_{\doubleunderline{\sigma}\doubleunderline{\sigma},ij} =& \sum_{T\in\mathcal{T}_h} - \int_T \frac{1}{2\nu}\phi_{\doubleunderline{\sigma},i}\,:\,\phi_{\doubleunderline{\sigma},j}\:d\underline{x}, \\
B_{\doubleunderline{\gamma}\doubleunderline{\sigma},ij} =& \sum_{T\in\mathcal{T}_h} - \int_T\phi_{\doubleunderline{\sigma},j}\,:\,\phi_{\doubleunderline{\gamma},i}\:d\underline{x}, \\
B_{\underline{u}\doubleunderline{\sigma},ij} =& \sum_{T\in\mathcal{T}_h} \bigg( -\int_T (\nabla\,\cdot\,\phi_{\doubleunderline{\sigma},j})\,\cdot\,\phi_{\underline{u},i}\:d\underline{x}
+ \int_{\partial T}  (\phi_{\doubleunderline{\sigma},j})_{nn} (\phi_{\underline{u},i}\,\cdot\,\underline{n})\:d\underline{s}  \bigg), \\
B_{\underline{\hat{u}}\doubleunderline{\sigma},ij} =& \sum_{T\in\mathcal{T}_h} \int_{\partial T}  (\phi_{\doubleunderline{\sigma},j})_{nt}\,\cdot\,\phi_{\underline{\hat{u}},i}\:d\underline{s}, \\
M_{\underline{u}\underline{u},ij} =& \sum_{T\in\mathcal{T}_h} \int_T \frac{1}{\Delta t}\phi_{\underline{u},i}\,\cdot\,\phi_{\underline{u},j}\:d\underline{x}, \\
C_{\underline{u}\underline{u},ij} =& \sum_{T\in\mathcal{T}_h}\int_T(\nabla\phi_{\underline{u},j}\cdot\phi_{\underline{u},j})\cdot\phi_{\underline{u},i}\:d\underline{x} 
+ \sum_{E\in\mathcal{F}_h^I} \int_{E} \big( -(\phi_{\underline{u},j}\,\cdot\,\underline{n})\llbracket\phi_{\underline{u},j}\rrbracket\,\cdot\,\langle\phi_{\underline{u},i}\rangle + \frac{1}{2}|\phi_{\underline{u},j}\,\cdot\,\underline{n}|\llbracket\phi_{\underline{u},j}\rrbracket\,\cdot\,\llbracket\phi_{\underline{u},i}\rrbracket \big) d\underline{s}.
\end{align*}
Due to the introduction of the additional variable $\underline{\hat{u}}_h$, the diagonal part of the system matrix for $\mathbf{\doubleunderline{\sigma}}$ and $\mathbf{\doubleunderline{\gamma}}$ is block diagonal. This projection problem is inf-sup stable and therefore
\begin{equation*}
M =
\begin{pmatrix}
M_{\doubleunderline{\sigma}\doubleunderline{\sigma}} & B_{\doubleunderline{\gamma}\doubleunderline{\sigma}}^t  \\
B_{\doubleunderline{\gamma}\doubleunderline{\sigma}} & 0
\end{pmatrix},
\end{equation*}
is invertible (see \cite{lederer3}).
By using static condensation of $\mathbf{\doubleunderline{\sigma}}$ and $\mathbf{\doubleunderline{\gamma}}$, the Schur complement reads as
\begin{align*}
S = A - BM^{-1}B^t,
% \end{equation*}
\quad 
\text{with}
\quad
% \begin{align*}
 A =
\begin{pmatrix}
M_{\underline{u}\underline{u}} & 0  \\
0 & 0
\end{pmatrix},
\quad 
\text{and}
\quad
 B =
\begin{pmatrix}
B_{\underline{u}\doubleunderline{\sigma}} & 0  \\
B_{\underline{\hat{u}}\doubleunderline{\sigma}} & 0
\end{pmatrix}.
\end{align*}
Note, that $-BM^{-1}B^t$ is spectral equivalent to an HDG-like stiffness matrix for $(\mathbf{\underline{u}^*}, {\underline{\hat{\mathbf{u}}}})$, and thus the Schur complement $S$, as a sum of the velocity mass matrix $A$ and $-BM^{-1}B^t$, is symmetric positive definite.

\begin{remark} \label{highervelbubbles} The base functions of $V_h$ can be divided into two types: face-based and cell-based, as described
in \cite{zaglmayr}.
The cell-based base functions ("element bubbles") are of higher order, with its support confined to an element $T\in\mathcal{T}_h$ and a vanishing normal trace on $\partial T$.
Since the support of these basis functions do not overlap the corresponding subpart of $M_{\underline{u}\underline{u}}$ is block diagonal and invertible.
Therefore, by applying static condensation to these high order velocity functions, a second Schur complement can be formulated, though this is not covered in this work. For more insight we refer to \cite{lederer3}.
\end{remark}

\begin{figure}

\centering
\input{./tikz/bddc.tex}
% \caption{Illustration of overlapping domain decomposition with (left) and without 
% (right) constraint. Red/blue arrows and circles indicate the respective edge/face-based
% degrees of freedom of two neighbouring Raviart-Thomas elements. Green arrow represents degrees of freedom
% of the global coarse problem. }
\caption{Illustration for $\mathcal{X}_h$ (left): Local basis functions (green circles) on two elements and the normal continuous (blue arrows) and tangential continuous (orange arrows) edge basis functions on the common edge. Illustration for $\mathcal{X}^{BDDC}_h$ (right): The normal/tangential continuous basis functions where "broken" in two parts. The solid arrows are associated to the left element, and the dashed ones to the right element. The operator $R$ averages the coefficients of corresponding solid and dashed basis functions.}

\label{DD}

\end{figure}
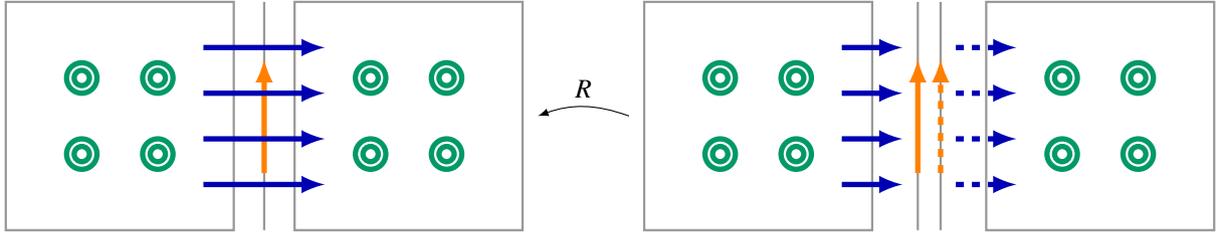

The global system then reads as
\begin{equation*}
S
\begin{pmatrix}
\mathbf{\underline{u}^*}  \\
{\underline{\hat{\mathbf{u}}}}
\end{pmatrix}
= 
\begin{pmatrix}
M_{\underline{u}\underline{u}}\mathbf{\underline{u}^{n}}-C_{\underline{u}\underline{u}}\mathbf{\underline{u}^{n}}  \\
0
\end{pmatrix}
,
\end{equation*}
which we are solving by means of a preconditioned conjugate gradient method. As preconditioner we make use of the balancing domain decomposition by constraints (BDDC) technique, see \cite{bddc}, where the domain decomposition is given by the triangulation (also called element-wise BDDC). 
% However, for our problem, we neglect the main idea of having "constraints", normally resulting in a low-order (e.g. smaller) global system needed to preserve the kernel of the considered finite element matrix. In contrast we consider solely local (solvable) systems. The main motivation why this is a feasible approach is that for small viscosities $\nu \ll 1$ and probably small time steps $\Delta t$ the mass matrix $A$ is the dominating factor when solving for $S$. 
The primary constraints of the BDDC lead to a global system. For small time steps and viscosities (i. e. $\nu \cdot \Delta t \ll 1$), the mass term $A$ in the definition of $S$ dominates. In this case we can skip the primary constraints, which results in a purely local solver.
The preconditioner can then be defined by considering an additional space $\mathcal{X}_h^{BDDC} \supset \mathcal{X}_h = V_h \times \hat{V}_h$, which is larger and has less conformity than $\mathcal{X}_h$. More precisely, we define $\mathcal{X}_h^{BDDC} = V_h^{disc.} \times \hat V_h^{disc.}$ where $\hat{V}_h^{disc.} = \{\hat{\underline{v}}_h \in \hat{V}_h(\partial T): \forall T \in \mathcal{T} \}$ with the local space 
\begin{align*}
	% \hat{V}_h^{disc.}=\{\hat{\underline{v}}_h\in \mathbb{P}^{k}(\partial \mathcal{T}_h, \mathbb{R}^3):\hat{\underline{v}}_h\,\cdot\,\underline{n}=0 \; \forall E\in \partial T, \forall T \in \mathcal{T}_h\}.
	% \hat{V}_h^{disc.} = \{\hat{\underline{v}}_h \in \hat{V}_h(\partial T): \forall T \in \mathcal{T} \} 
	% \quad 
	% \text{with the local space}
	% \quad
	\hat{V}_h(\partial T)=\{\hat{\underline{v}}_h\in L^2(\partial T, \mathbb{R}^3):\hat{\underline{v}}_h\big|_E \in \mathbb{P}^k(E, \mathbb{R}^3),\hat{\underline{v}}_h\,\cdot\,\underline{n}=0 \; \forall E\in \partial T\}.
\end{align*}
Note that in contrast to $\hat{V}_h$, functions in $\hat{V}_h^{disc.}$ are not single but dual valued on facets $E \in \mathcal{F}_h$. Thus, compared to $\mathcal{X}_h$, the space $\mathcal{X}_h^{BDDC}$ "breaks" the normal/tangential continuity of basis functions in $\mathcal{X}_h$ which are associated to facets, see Figure~\ref{DD} for a visualization.
Now let $R:\mathcal{X}_h^{BDDC} \to \mathcal{X}_h$ be the averaging operator that averages the entries of the coefficient vector associated to the aforementioned "broken" facet basis functions to derive a function in $\mathcal{X}_h$. Further let $S^{BDDC}$ be the block matrix that results from assembling the bilinear-form associated to the matrix $S$ but on the "broken" space $\mathcal{X}_h^{BDDC}$ on each element separately, see also Remark~\ref{rem::bddcassembly}. Then we define our solver by 
\begin{equation*}
	C^{-1}_{BDDC} = R S_{BDDC}^{-1} R^t.
\end{equation*}
Due to its block structure, solving for $S_{BDDC}^{-1}$ allows for optimal parallel scalability. Further we want to emphasize that due to the application of $R$ the solution (when solving with $C^{-1}_{BDDC}$) is in $\mathcal{X}_h$, i.e. provides the proper normal/tangential continuity.

\begin{remark} \label{rem::bddcassembly}
	In practice the space $\mathcal{X}_h^{BDDC}$ is neither implemented nor constructed. The computation of $S_{BDDC}^{-1}$ occurs during the assembly of $S$ where the local element-wise contributions of $S_{BDDC}$ are inherently calculated.
\end{remark}

For the second step, the hybrid mixed problem from \eqref{MCS_step2} can be written in matrix form as
\begin{equation}
\begin{pmatrix}
M_{\underline{\tilde{u}}\underline{\tilde{u}}} & B_{p\underline{\tilde{u}}}^t & B_{\hat{p}\underline{\tilde{u}}}^t \\
B_{p\underline{\tilde{u}}} & 0 & 0 \\
B_{\hat{p}\underline{\tilde{u}}} & 0 & 0
\end{pmatrix}
\begin{pmatrix}
\mathbf{\underline{\tilde{u}}}  \\
\mathbf{p}  \\
\mathbf{\hat{p}}
\end{pmatrix}
=
\begin{pmatrix}
0  \\
B_{p\underline{u}}\mathbf{\underline{u}^*}  \\
0
\end{pmatrix},
\label{hmp}
\end{equation}
where
\begin{align*}
& M_{\underline{\tilde{u}}\underline{\tilde{u}},ij} = \sum_{T\in\mathcal{T}_h} - \int_T \phi_{\underline{\tilde{u}},i}\,\cdot\,\phi_{\underline{\tilde{u}},j}\:d\underline{x}, \\
& B_{p\underline{\tilde{u}},ij} = \sum_{T\in\mathcal{T}_h} \int_T (\nabla\,\cdot\,\phi_{\underline{\tilde{u}},j})\phi_{p,i}\:d\underline{x}, \\
& B_{\hat{p}\underline{\tilde{u}},ij} = \sum_{T\in\mathcal{T}_h} \int_{\partial T} (\phi_{\underline{\tilde{u}},j}\,\cdot\,\underline{n})\phi_{\hat{p},i}\:d\underline{s}, \\
& B_{p\underline{u},ij} = \sum_{T\in\mathcal{T}_h} \int_T (\nabla\,\cdot\,\phi_{\underline{u},j})\phi_{p,i}\:d\underline{x}.
\end{align*}
With the introduction of $\hat{p}_h$, the DOFs associated with $\tilde{u}_h$ and $p_h$ consist of local problems and only couple with base functions of $\hat{q}_h$. In the same way as before, static condensation allows us to decompose the system matrix from \eqref{hmp} and significantly reduce the size of the final problem that needs to be solved.
Similar as before, the Schur complement for this mixed problem
is 
\begin{align*}
S_p = - B_pM_p^{-1}B_p^t,
\quad
\text{with}
\quad
 M_p =
\begin{pmatrix}
M_{\underline{\tilde{u}}\underline{\tilde{u}}} & B_{p\underline{\tilde{u}}}^t  \\
B_{p\underline{\tilde{u}}} & 0 
\end{pmatrix},
\quad 
\text{and}
\quad
 B_p =
\begin{pmatrix}
B_{\hat{p}\underline{\tilde{u}}}  & 0
\end{pmatrix}.
\end{align*}
The matrix $S_p$ is symmetric and positive definite, see also Section 7.2 in \cite{boffi2013mixed}, \xaver{and can be interpreted as a pressure stiffness
matrix}. To this end, to solve this system we use the algebraic multigrid preconditioner GAMG from PETSc's native AMG framework \cite{PETSC}.

%% file: tikz/bddc.tex
% Author: Izaak Neutelings (November 2020)
% \documentclass[convert={density=300,size=1080x800,outext=.png}]{standalone}
% \usepackage{tikz}
% \usetikzlibrary{arrows.meta}
% \usetikzlibrary{calc}
% \tikzset{>=latex}

\colorlet{myelcol}{black!40!white}
\colorlet{mydarkblue}{blue!40!black}
\colorlet{myblue}{blue!70!black}
\colorlet{mygreen}{green!60!blue}
\colorlet{myred}{red!65!black}
\colorlet{myorange}{orange!90!black!90}

% \begin{document}

\begin{tikzpicture}
    \def\L{3}
    \def\N{4}
    \def\NN{2}
    \def\len{ 0.4}
    \def\d{0.8}
    \def\dd{0.6}
    \def\veclen{1.5}

    \draw[line width = 0.3mm, color=myelcol] (0,0) rectangle (\L, \L);
    \begin{scope}[xshift=\L cm + \d cm]
        \draw[line width = 0.3mm, color=myelcol] (0,0) rectangle (\L, \L);
        \foreach \x in {1, ..., \NN} 
        {
            \foreach \y in {1, ..., \NN} 
            {
            \pgfmathsetmacro\h{ \y/(\NN+1) * \L} 
            \pgfmathsetmacro\h{ \y/(\NN+1) * \L} 
            \draw [->, line width = 0.70mm, color=mygreen, variable=\y] (\x, \y) circle (0.2);
            \draw [->, line width = 0.70mm, color=mygreen, variable=\y] (\x, \y) circle (0.1);
            }
        }
    \end{scope}
    
    \draw[line width = 0.3mm, color=myelcol] (\L + \d/2,0) to (\L + \d/2,\L);
    % \draw[line width = 0.3mm, color=myelcol] (- \d/2,0) to (- \d/2,\L);
    % \draw[line width = 0.3mm, color=myelcol] (0, - \d/2) to (\L, - \d/2);
    % \draw[line width = 0.3mm, color=myelcol] (0, \L + \d/2) to (\L, \L + \d/2);

    % \draw[line width = 0.3mm, color=myelcol] (0, - \d) to (\L, - \d);
    % \draw[line width = 0.3mm, color=myelcol] (0, - \d) to (0, - \d*1.5);
    % \draw[line width = 0.3mm, color=myelcol] (\L, - \d) to (\L, - \d*1.5);

    \draw[->, line width = 0.70mm, color = orange] (\L + \d/2,\L/2-\veclen/2) to (\L + \d/2,\L/2+\veclen/2);
    
    \foreach \x in {1, ..., \NN} 
    {
        \foreach \y in {1, ..., \NN} 
        {
        \pgfmathsetmacro\h{ \y/(\NN+1) * \L} 
        \pgfmathsetmacro\h{ \y/(\NN+1) * \L} 
        \draw [->, line width = 0.70mm, color=mygreen, variable=\y] (\x, \y) circle (0.2);
        \draw [->, line width = 0.70mm, color=mygreen, variable=\y] (\x, \y) circle (0.1);
        }
    }

     \foreach \y in {1, ..., \N} 
    {
       \pgfmathsetmacro\h{ \y/(\N+1) * \L} 
       \draw [->, line width = 0.70mm, color=myblue, variable=\y] (\L-\len, \h) -- (\L+\d+\len, \h);
    %    \draw [->, line width = 0.70mm, color=myblue, variable=\y] (\len, \h) -- (-\len-\d, \h);
     }

    %  \foreach \y in {1, ..., \N} 
    % {
    %    \pgfmathsetmacro\h{ \y/(\N+1) * \L} 
    %    \draw [->, line width = 0.70mm, color=myblue, variable=\y] (\h, \len) -- (\h, -\len-\d);
    %    \draw [->, line width = 0.70mm, color=myblue, variable=\y] (\h, \L -\len) -- (\h, \L +\len);
    %  }

    \draw[<-, bend left = 20] (2*\L+\d + 0.2, \L/2) to node[midway, above] {$R$}  (2.8*\L - 0.2, \L/2) ;

     \begin{scope}[xshift=2.8*\L cm]
        \draw[line width = 0.3mm, color=myelcol] (0,0) rectangle (\L, \L);
        \begin{scope}[xshift=\L cm + 2.5*\dd cm]
            \draw[line width = 0.3mm, color=myelcol] (0,0) rectangle (\L, \L);
            \foreach \x in {1, ..., \NN} 
            {
                \foreach \y in {1, ..., \NN} 
                {
                \pgfmathsetmacro\h{ \y/(\NN+1) * \L} 
                \pgfmathsetmacro\h{ \y/(\NN+1) * \L} 
                \draw [->, line width = 0.70mm, color=mygreen, variable=\y] (\x, \y) circle (0.2);
                \draw [->, line width = 0.70mm, color=mygreen, variable=\y] (\x, \y) circle (0.1);
                }
            }
        \end{scope}
        \draw[line width = 0.3mm, color=myelcol] (\L + \dd,0) to (\L + \dd,\L);
        \draw[line width = 0.3mm, color=myelcol] (\L + 1.5*\dd,0) to (\L + 1.5* \dd,\L);
        
        \draw[->, line width = 0.70mm, color = orange] (\L + \dd,\L/2-\veclen/2) to (\L + \dd,\L/2+\veclen/2);
        \draw[->, line width = 0.70mm, dashed, color = orange] (\L + \dd*1.5,\L/2-\veclen/2) to (\L + \dd*1.5,\L/2+\veclen/2);

        \foreach \x in {1, ..., \NN} 
        {
            \foreach \y in {1, ..., \NN} 
            {
            \pgfmathsetmacro\h{ \y/(\NN+1) * \L} 
            \pgfmathsetmacro\h{ \y/(\NN+1) * \L} 
            \draw [->, line width = 0.70mm, color=mygreen, variable=\y] (\x, \y) circle (0.2);
            \draw [->, line width = 0.70mm, color=mygreen, variable=\y] (\x, \y) circle (0.1);
            }
        }

        \foreach \y in {1, ..., \N} 
        {
        \pgfmathsetmacro\h{ \y/(\N+1) * \L} 
        \draw [->, line width = 0.70mm, color=myblue, variable=\y] (\L-\len, \h) -- (\L+\len, \h);
        \draw [->, line width = 0.70mm, dashed, color=myblue, variable=\y] (\L-\len +2.5*\dd , \h) -- (\L+\len+2.5*\dd, \h);
        }
     \end{scope}

\end{tikzpicture}
% \end{document}

%% file: chapters/hopu.tex
\section{Methodology}\label{methods}

In the following, we introduce two methodologies explored in this study.
First, we provide a brief overview of the standard ILES approach, which does not use an explicit turbulence model. Next, we outline the HOPU
method, which enhances the inherent dissipation mechanism of ILES. Explicit turbulence modeling approaches have been excluded from this study
due to their inferior performance compared to ILES in a DG framework for wall-bounded flow problems (see e.g., \cite{HOPU2023,Peraire2017}). 
 
\subsection{ILES} 

\xaver{Unlike explicit LES schemes,
ILES utilizes the discretization of the governing equations to naturally provide dissipation at
small-resolved scales.} In the discrete model 
\eqref{MCS}, \xaver{we employ an upwind
Riemann solver to strengthen robustness and control dissipation
on the small-scale structures.}
The stability characteristics of DG methods make them particularly suitable for multiscale
phenomena problems, as dissipation is inherently applied to the high-order
part of the solution only. This capability was validated for the linear convection-diffusion equation in \cite{moura2}, affirming the
effectiveness of DG formulations for simulating under-resolved turbulent flow problems.
In this work, we refer to ILES, when the standard formulation of Section \ref{mcs} is taken. 
 
\subsection{High-order projection-based upwind ILES}\label{hopu_chapter}

In simulations of under-resolved turbulent flows via the ILES approach, unresolved eddies are taken into account inherently via an additional stabilization by (tangential) velocity jumps at element interfaces due to the upwind Riemann solver. The latter can be interpreted as an additional numerical diffusion with an implicitly defined "eddy viscosity". However, the jump terms penalize, i.e. dampen, the entire range of resolved scales of the velocity solution. Similar to variational multiscale simulation (VMS) approaches (see \cite{Hughes2000, VMS}) the primary concept of the high-order  projection-based upwind (HOPU) method is to separate the resolved range of scales. Here, numerical dissipation exclusively acts on the small-resolved structures (those approximated by the high-order component of the velocity approximation).
Consequently, the low-frequent subrange of the velocity jump is considered in a central flux manner. The separation of scales in the jumps is achieved through an additional $L^2$-projection. Therefore, we define the space 
\begin{equation*}
\hat{V}_h^{proj}=\{\underline{\hat{v}}_h^{proj}\in \mathbb{P}^l(\mathcal{F}_h, \mathbb{R}^3):\underline{\hat{v}}_h^{proj}\,\cdot\,\underline{n}=0 \; \text{on} \; E\in\mathcal{F}_h\},
\end{equation*} 
where the polynomial order is chosen such that  $0 \leq l \leq k$.
Then, the convective term is changed to 
\begin{align} \label{eq::hopudefinition}
    c^{HOPU}_h(\underline{u}_h,\underline{u}_h,\underline{v}_h)=  &\sum_{T\in\mathcal{T}_h}\int_T(\nabla\underline{u}_h\cdot\underline{u}_h)\cdot\underline{v}_h\:d\underline{x}  \\
     &+\sum_{E\in\mathcal{F}_h^I} \bigg(- \int_{E}(\underline{u}_h\,\cdot\,\underline{n})\llbracket\underline{u}_h\rrbracket\,\cdot\,\langle\underline{v}_h\rangle\:d\underline{s} 
  +\int_{E}\frac{1}{2}|\underline{u}_h\,\cdot\,\underline{n}| (I-\Pi^{l}_{\hat{V}})(\llbracket\underline{u}_h\rrbracket)\,\cdot\,(I-\Pi^{l}_{\hat{V}})(\llbracket\underline{v}_h\rrbracket)\:d\underline{s}\Big),
\end{align}
where $\Pi^{l}_{\hat{V}}: L^2(E) \rightarrow
\hat{V}_h^{proj}(E)$ is the facet wise $L^2$-projection into $\hat{V}_h^{proj}(E)$.
The idea is to be able to control the spectrum on which the diffusion, introduced via the last integral in \eqref{eq::hopudefinition}, is applied on. For $l=k$, we have $\Pi^{l}_{\hat{V}} = I$. Consequently, no upwinding occurs, resulting in a central treatment of the convection where no stabilization via numerical diffusion is applied at all (for all scales). For $0 \le l < k$, the operator $(I-\Pi^{l}_{\hat{V}})$ removes any stabilization that is applied on the lower frequencies, i.e. the lower order part of the solution. The range of scales can accordingly be adjusted by the choice of $l$. Eventually, when $\Pi^{l}_{\hat{V}}$ is chosen as the "zero-operator" we retrieve the standard upwinding from \eqref{upwind}. For a comment on the stability and implementation of HOPU we refer to \cite{HOPU2023}.

A way to address the problem of determining the local polynomial order used in $\Pi^l_{\hat{V}}$ was introduced by the authors in \cite{HOPU2023}. This approach defines a parameter to estimate
the magnitude of the jumps across element interfaces. The average absolute value jump is given by
\begin{equation*}
\eta = \frac{\int_{E} |(I-\Pi^{l}_{\hat{V}})(\llbracket\underline{u}_h\rrbracket) |\:d\underline{s}}
{\int_{E} | \langle\underline{u}_h\rangle |\:d\underline{s}}.
\end{equation*}

Based on this local (facet-wise) $\eta$, we determine the polynomial order used. We define values $\eta_i$ such that
$\eta_{-1}=0 < \eta_0 < \eta_1 < ... < \eta_{k+1}=1$, which are problem-specific thresholds for the adaptive HOPU method.
Therefore, for $\eta_i < \eta_{loc} \le \eta_{i+1}$ we use the local order $l_{loc}=i$ only when
$i \ge 0$.
In the case of $\eta < \eta_0$ the adaptive algorithm reverts back to the standard convection stabilization from 
\eqref{upwind}.
In our simulations the local polynomial order $l_{loc}$ was
determined after each 10th time step and adjusted accordingly.

%% file: chapters/results.tex
\section{Eppler 387 wing}\label{results}

In this Section, we compare the accuracy of high-order ILES in predicting transition and turbulent boundary layer with that of the HOPU method. Additionally, we compare our results to the experimental data provided in \cite{Nasa1988}, if available.
We consider two distinct cases of transitional flow over the Eppler 387 wing.
Firstly, we examine a flow scenario at Reynolds numbers $Re = 1 \cdot 10^5$ and an angle of attack of $\alpha= 10\degree$.
The second case involves a higher Reynolds number of $Re = 3 \cdot 10^5$ with a reduced angle of attack set to $\alpha = 4\degree$.
In the following, we use the abbreviations $R1/\alpha10$ and $R3/\alpha4$ for the two respective flow scenarios.
Both cases exhibit laminar boundary layer separation and subsequent turbulent reattachment on the suction side of the airfoil.
In the case of a high angle of attack $R1/\alpha10$, a nearly fully turbulent boundary layer can be observed across the entire upper surface of the airfoil. In contrast, despite the higher Reynolds number,  in the case $R3/\alpha4$ separation only occurs at approximately $x/c \approx 0.5$. These configurations were selected to analyse the behaviour of HOPU to ILES under different flow scenarios. With regards to \cite{HOPU2023}, where an improvement using HOPU was observed in wall-bounded turbulence for channel flows, we particularly expect an improvement in cases with a wider area of turbulence, i.e. the case $R1/\alpha10$. 

\begin{figure}
\centering
	\begin{subfigure}[t]{0.4\textwidth}
	\centering
    	\includegraphics[width=\textwidth]{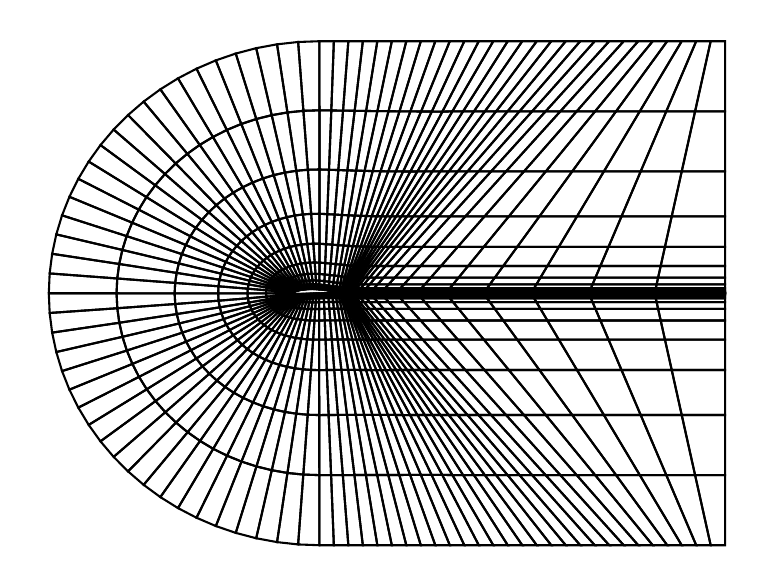}
		\caption{}
		\centering
		\label{fig:mesh}
	\end{subfigure}
~
	\begin{subfigure}[t]{0.4\textwidth}
		\centering
    	\includegraphics[width=0.75\textwidth]{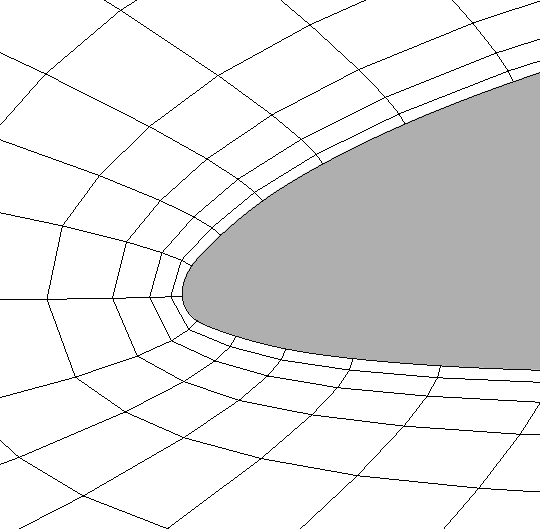}
		\caption{´}
		\label{fig:curved}
	\end{subfigure}
	\caption{(a) computational mesh, (b) curved boundary elements.}
	\label{fig:mesh_curved}
\end{figure}

To accommodate the inherently three-dimensional nature of turbulent flows, a 3D mesh is constructed by initially defining a two-dimensional curved quadrilateral mesh and subsequently extruding it in the spanwise direction. Figure \ref{fig:mesh_curved} shows the obtained \xaver{2D} grid.

A given inflow velocity is imposed on the half circle (inlet), while homogeneous Neumann boundary conditions are applied on the border of the box (outlet). No-slip conditions are enforced at the airfoil and periodic boundary conditions are set for the remaining faces normal to the $z$-axis. 
As effects due to the finite extent of the computational domain are assumed to be minor, the inlet flow angle is not
corrected and no freestream turbulence intensity has been introduced at the inlet.
By defining $c$ the chord length of the profile, the radius and
box length were chosen to $r = 5c$ and $b = 7c$, respectively. The extrusion length of the mesh is set to $d = 0.1c$.
The details of the different meshes are summarized in Table \ref{table1}. The polynomial degree of the MCS discretization is set to $k = 3$ and the time steps
were chosen to be $\Delta t = 5 \cdot 10^{-4}$ for $R3/\alpha4$ and $\Delta t = 3 \cdot 10^{-4}$ for $R1/\alpha10$. A fully converged (in a statistical stationary sense) flow achieved from a coarser resolution has
been used as initial conditions for all presented simulation runs. Averaging starts after 5 characteristic time
lengths ($U_{\infty}/c$) and was stopped after 10 full characteristic time lengths have been reached. Velocity, pressure and stress
samples are taken every 10th time step. The relevant quantities of interest are the 
three velocity components $\underline{u}_h = (u, v, w)^t$ and the pressure 
$p_h$, where the $h$-index is skipped for readability. The mean operator
$\overline{(*)}$ involves averaging over time and
in the homogenous $z$-direction.

\begin{table}
\begin{center}
\begin{tabular}{ |c|c|c|c|c|c| } 
\hline
Mesh & Hexahedrons & $\Delta y_w/c * 10^{3}$ & $DOFs * 10^{-6}$ (M) & $DOFs* 10^{-6}$ (P) & $gDOFs* 10^{-3}$ (P) \\
\hline
m1 & 2352 & 2.6 & 1.980 & 0.828 & 112.8 \\ 
m2 & 7200 & 1.6 & 6.062 & 2.534 & 345.5 \\ 
m3 & 14400 & 1.2 & 12.12 & 5.069 & 691.2 \\ 
\hline
\end{tabular}
\end{center}
\caption{Details of the computational meshes considered for the Eppler 387 wing. Here, $\Delta y_w$ is the wall-nearest cell height and gDOFs indicates the number of unknowns of the condensed system. 
We refer to the momentum equation \eqref{MCS_step1} as (M) and the projection defined in \eqref{MCS_step2} as (P).}
\label{table1}
\end{table}

The finest mesh was selected such that the solution of the pressure coefficient is grid
independent, thereby establishing its results as a reference for comparison across different resolutions.
Grid convergence is considered as the $L^2$-norm of the
difference of mean pressure coefficient between consecutive meshes is below
$0.01$. This criterion is achieved for the finest mesh m3. Mesh m1 and m2 were chosen
with coarser resolutions specifically
to study highly to moderately under-resolved configurations.

\begin{figure}
\centering
\begin{subfigure}{0.8\textwidth}
    \includegraphics[width=\textwidth]{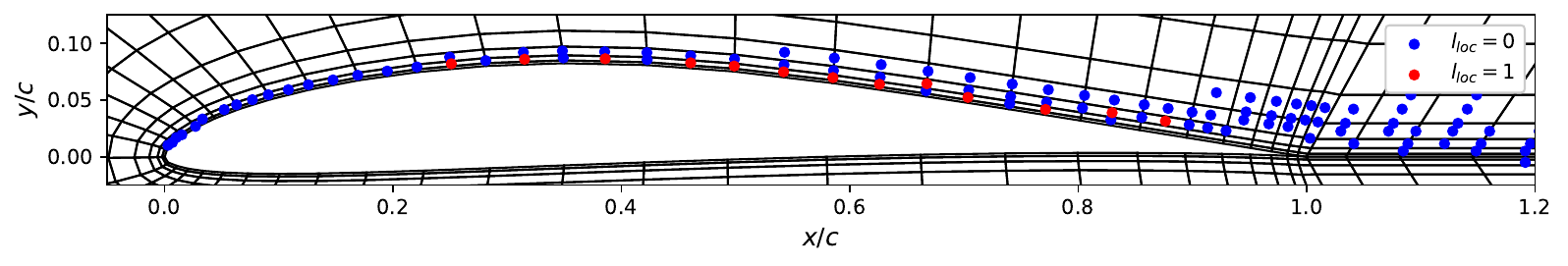}
    \caption{}
    \label{fig:l_loc_a10}
\end{subfigure}
\hfill
\begin{subfigure}{0.8\textwidth}
    \includegraphics[width=\textwidth]{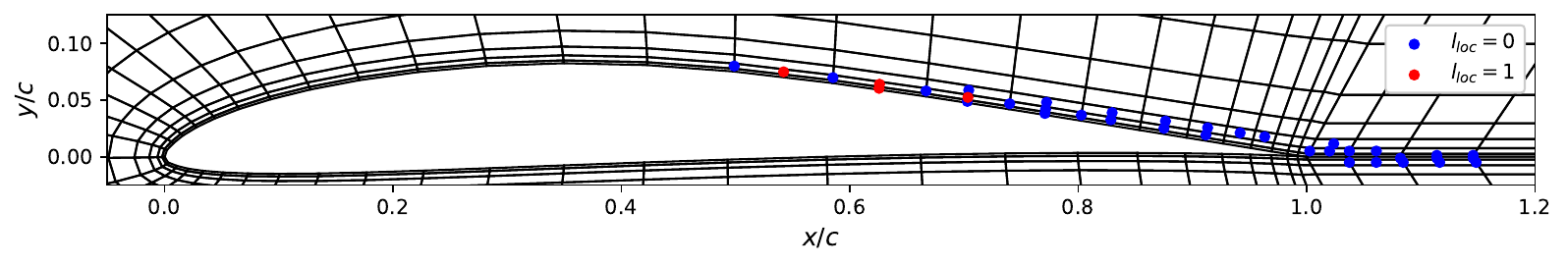}
	\caption{}
	\centering
	\label{fig:l_loc_a4}
\end{subfigure}    
\caption{Mode of local polynomial order $l_{loc}$ located in the middle point of 
its respective facet for (a) $R1/\alpha 10$ and (b) $R3/\alpha 4$.}
\label{fig:l_loc}
\end{figure}

For HOPU we have set the limits to $\eta = \{0.1,0.2,0.3,0.4\}$.
In the context of transient flow around an airfoil, it is observed that the magnitude of velocity jumps
tends to be significantly larger in the turbulent wall-bounded regions compared to the laminar and outer regions.
Figure \ref{fig:l_loc} shows the mode of the local polynomial order $l_{loc}$ over all time steps and
respective facets in the $z$-direction used by the HOPU approach with mesh m1. 

\subsection{Comparison of pressure and aerodynamic forces}

\begin{figure}
\centering
\begin{subfigure}{0.4\textwidth}
    \includegraphics[width=\textwidth]{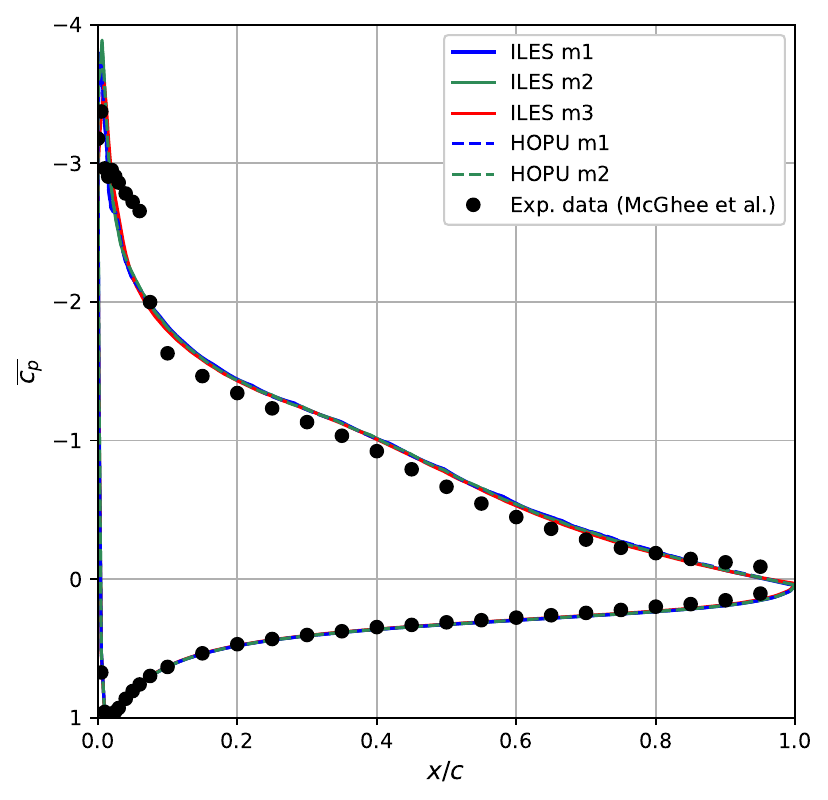}
    \caption{}
    \label{fig:cp_a10}
\end{subfigure}
~
\begin{subfigure}{0.4\textwidth}
    \includegraphics[width=\textwidth]{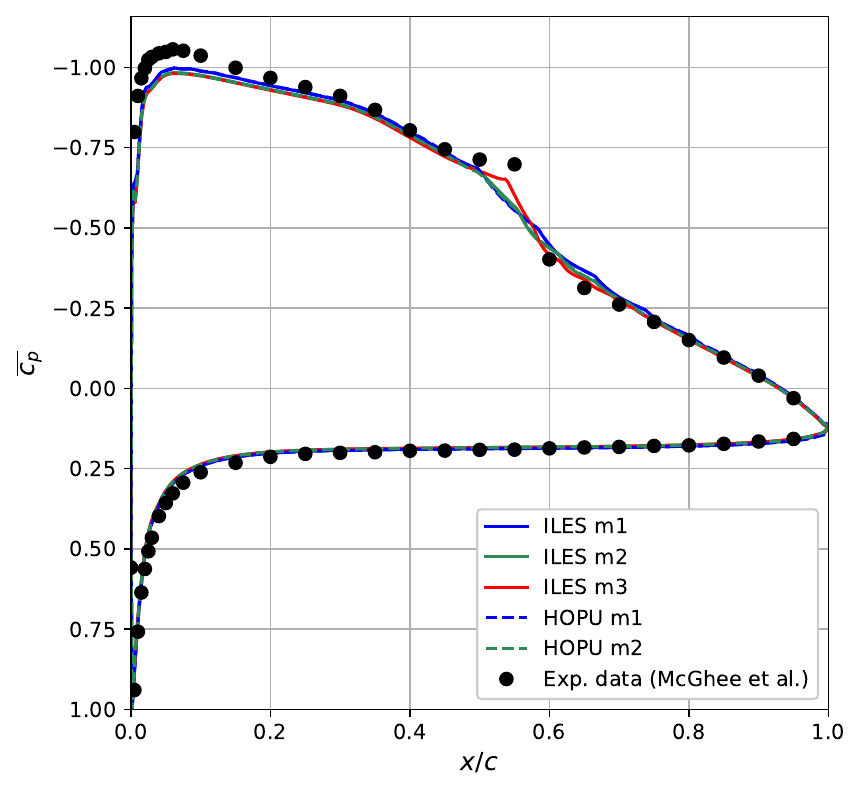}
	\caption{}
	\centering
	\label{fig:cp_a4}
\end{subfigure}    
\caption{Distribution of the pressure coefficient $\overline{c_p}$ over the Eppler 387 airfoil for (a) $R1/\alpha 10$ and (b) $R3/\alpha 4$
case.}
\label{fig:cp}
\end{figure}

Figure \ref{fig:cp} depicts the mean pressure coefficient computed
using ILES and HOPU. In general, the numerical results exhibit very good agreement
with the experimental data reported in \cite{Nasa1988},
except in the regions near the transition location (end of laminar separation bubble and 
subsequent pressure rise)
and close to the leading edge of the airfoil. For $R3/\alpha4$, a small pressure
plateau is caused by the separation bubble,
which is accurately predicted by the finest mesh m3 and aligns well with the experimental data.
In contrast, for $R1/\alpha10$, as the angle of attack
increases, the laminar separation bubble moves forward, resulting in significantly reduced recirculation area.  
The coarse meshes are not able to predict a meaningful bubble
leading to smaller separation areas or even an (unphysical) almost natural transition.
The mismatch near the leading edge concerning the 
experimental data primarily stems from to the finite extent of the
computational domain, which causes an underestimation of the pressure in this
area. 

Table \ref{table2} presents the lift and drag coefficients, along with the approximated
separation and reattachment locations of the present computations compared to experimental results.
Despite minor differences, all values of $c_l$ and $c_d$ closely match the experimental data.
The spatial extent of the separation bubble is characterized by $x_s/c$ and $x_r/c$.
It is observed that with increasing numerical resolution and accuracy of the method, the size of the recirculating flow region, represented by
$x_s/c$ and $x_r/c$, expands.
Specifically, for the finest mesh (ILES m3) compared to the experimental data from \cite{Nasa1988} for $R3/\alpha4$,
the experiment indicates an earlier flow separation than the numerical simulation.
We want to emphasize that no significant differences are observed between the HOPU method and ILES. This is a crucial observation since the goal of HOPU was to achieve a better approximation of turbulent flows (see Section \ref{sec::tubulentboundarylayer} below) without affecting key quantities, such as the pressure, drag and lift coefficients.
As these factors primarily rely on the pressure distribution and the separation and reattachment of the boundary layer, we do not anticipate a significant influence from the HOPU scheme.

%The mean wall shear stress distribution along the upper side of the wing is
%displayed in Figure \ref{fig:cp} (right). Good agreement of all numerical
%results is given in the laminar portion of the flow up to $x/c \approx 0.4$.
%This shows that even for the highly under-resolved case with mesh M1 
%no additional dissipation is introduced by the high-order
%ILES method in the laminar region. 
%Major discrepancies can be observed as separation and
%transition dominate the behaviour of the boundary layer flow. Results 
%originated by mesh M1 and M2 predicts only a small recirculation
%bubble and underestimate the stress in the reattached
%flow region. The
%separation and reattachment location, $x_s/c$ and $x_r/c$ respectively,
%are shown in Table \ref{table2}. The HOPU variant with mesh M1 predicts slightly
%higher wall shear stresses in the reattached turbulent region compared
%to ILES , as its 
%less dissipative mechanism allows in general for 
%higher gradients at the wall.

\begin{table}
\begin{center}
\begin{tabular}{ |c|c|c|c|c| } 
\hline
$R1/\alpha 10$ case & $c_l$ & $c_d$ & $x_s/c$ & $x_r/c$ \\
\hline
Exp. Data (McGhee et al.) & 1.200 & 0.0413 & - & - \\ 
ILES M1 & 1.274 & 0.0432 & 0.007 & 0.010 \\ 
HOPU M1 & 1.271 & 0.0436 & 0.007 & 0.010 \\ 
ILES M2 & 1.268 & 0.0421 & 0.008 & 0.016 \\ 
HOPU M2 & 1.266 & 0.0425 & 0.008 & 0.016 \\ 
ILES M3 & 1.259 & 0.0411 & 0.010 & 0.027 \\ 
\hline
$R3/\alpha 4$ case & $c_l$ & $c_d$ & $x_s/c$ & $x_r/c$ \\
\hline
Exp. Data (McGhee et al.) & 0.792 & 0.0109 & 0.40 & 0.58 \\ 
ILES M1 & 0.767 & 0.0116 & 0.498 & 0.517 \\ 
HOPU M1 & 0.767 & 0.0116 & 0.499 & 0.517 \\ 
ILES M2 & 0.763 & 0.0113 & 0.498 & 0.515 \\ 
HOPU M2 & 0.763 & 0.0113 & 0.498 & 0.515 \\ 
ILES M3 & 0.759 & 0.0113 & 0.467 & 0.560 \\ 
\hline
\end{tabular}
\end{center}
\caption{Lift and drag coefficient, separation and reattachment location for different cases.}
\label{table2}
\end{table}

\subsection{Comparison of turbulent boundary layer data} \label{sec::tubulentboundarylayer}

\begin{figure}
\centering
\begin{subfigure}{0.8\textwidth}
    \includegraphics[width=\textwidth]{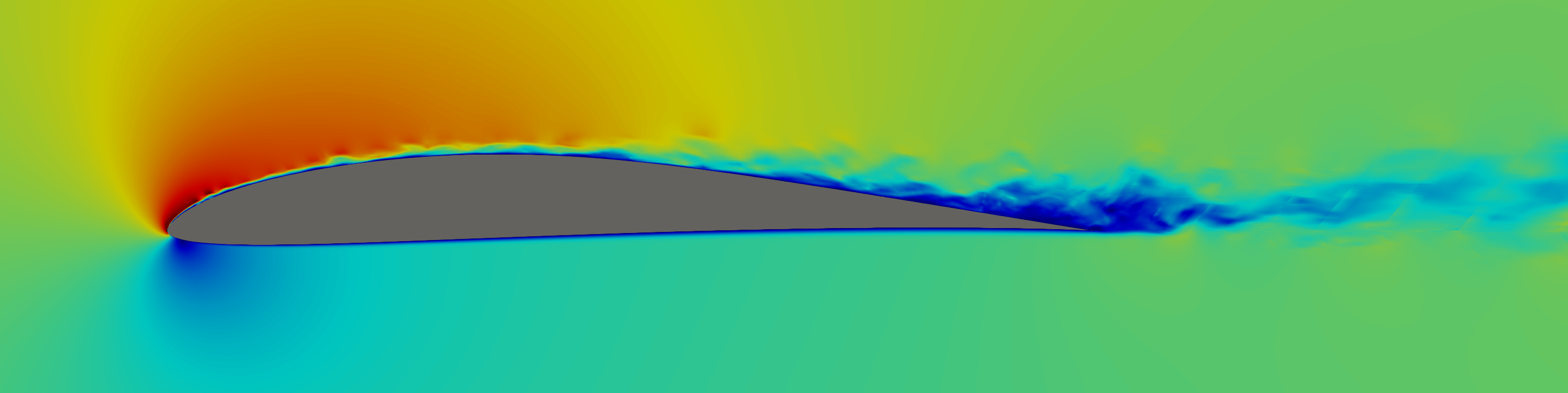}
    \caption{}
    \label{fig:instantvel_a10}
\end{subfigure}
\hfill
\begin{subfigure}{0.8\textwidth}
    \includegraphics[width=\textwidth]{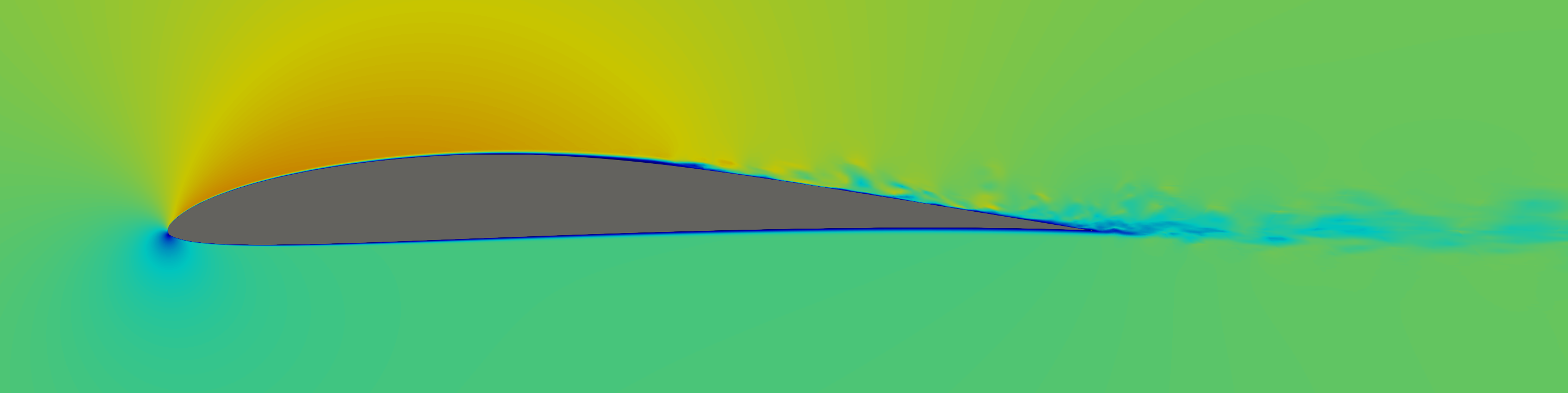}
	\caption{}
	\centering
	\label{fig:instantvel_a4}
\end{subfigure}    
\caption{Instantaneous velocity magnitude $|\underline{u}|/U_{\infty}$ field at given point in time for (a) $R1/\alpha 10$ and (b) $R3/\alpha 4$.}
\label{fig:instantvel}
\end{figure}

\begin{figure}
\centering
\includegraphics[width=0.8\textwidth]{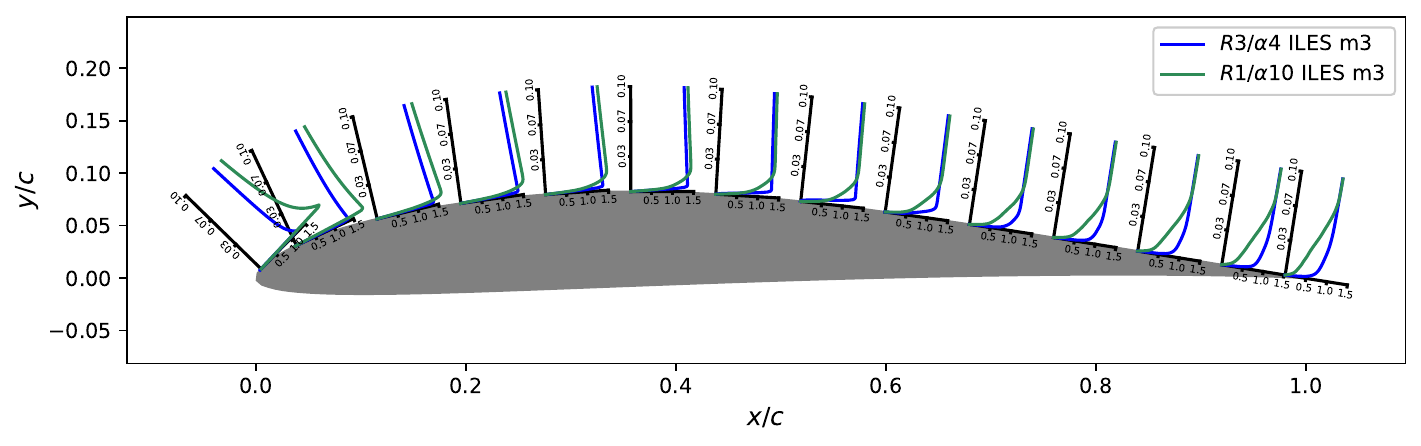}
\caption{Mean tangential velocity magnitude $|\overline{\underline{u}}_t|/U_{\infty}$ profiles along the upper surface of the Eppler 387 airfoil.}
\label{fig:uprofiles}
\end{figure}

By design, the Eppler 387 wing maintains a fully laminar boundary layer on its
lower surface despite the high Reynolds number. However, as
discussed previously, transition to turbulent flow occurs along the suction side,
as depicted in Figure \ref{fig:cp}.  The turbulent mixing characteristics lead to
rapid reattachment, and the subsequent turbulent boundary layer remains attached
all the way to the trailing edge. This specific behaviour of the boundary layer
is visually demonstrated in Figure \ref{fig:instantvel}
through the instantaneous velocity magnitude field. Furthermore, the mean
tangential velocity $|\overline{\underline{u}}_t|/U_{\infty}$ is computed at
various locations and plotted in Figure
\ref{fig:uprofiles}. Changes in the outer layer profiles can be observed as
the flow transitions into the turbulent regime.

\begin{figure}
\centering
\begin{subfigure}{0.4\textwidth}
    \includegraphics[width=\textwidth]{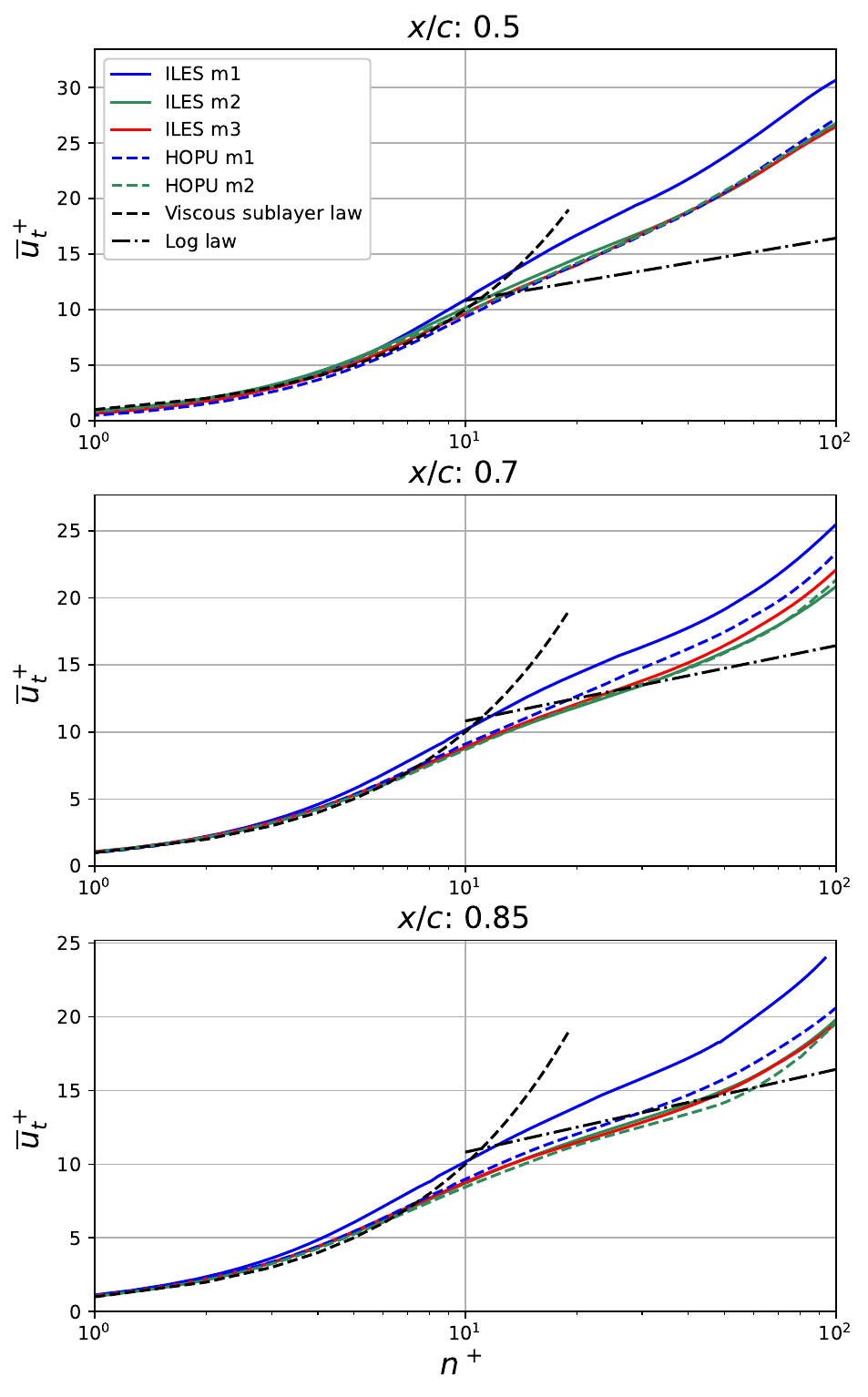}
    \caption{}
    \label{fig:ut_a10}
\end{subfigure}
~
\begin{subfigure}{0.4\textwidth}
    \includegraphics[width=\textwidth]{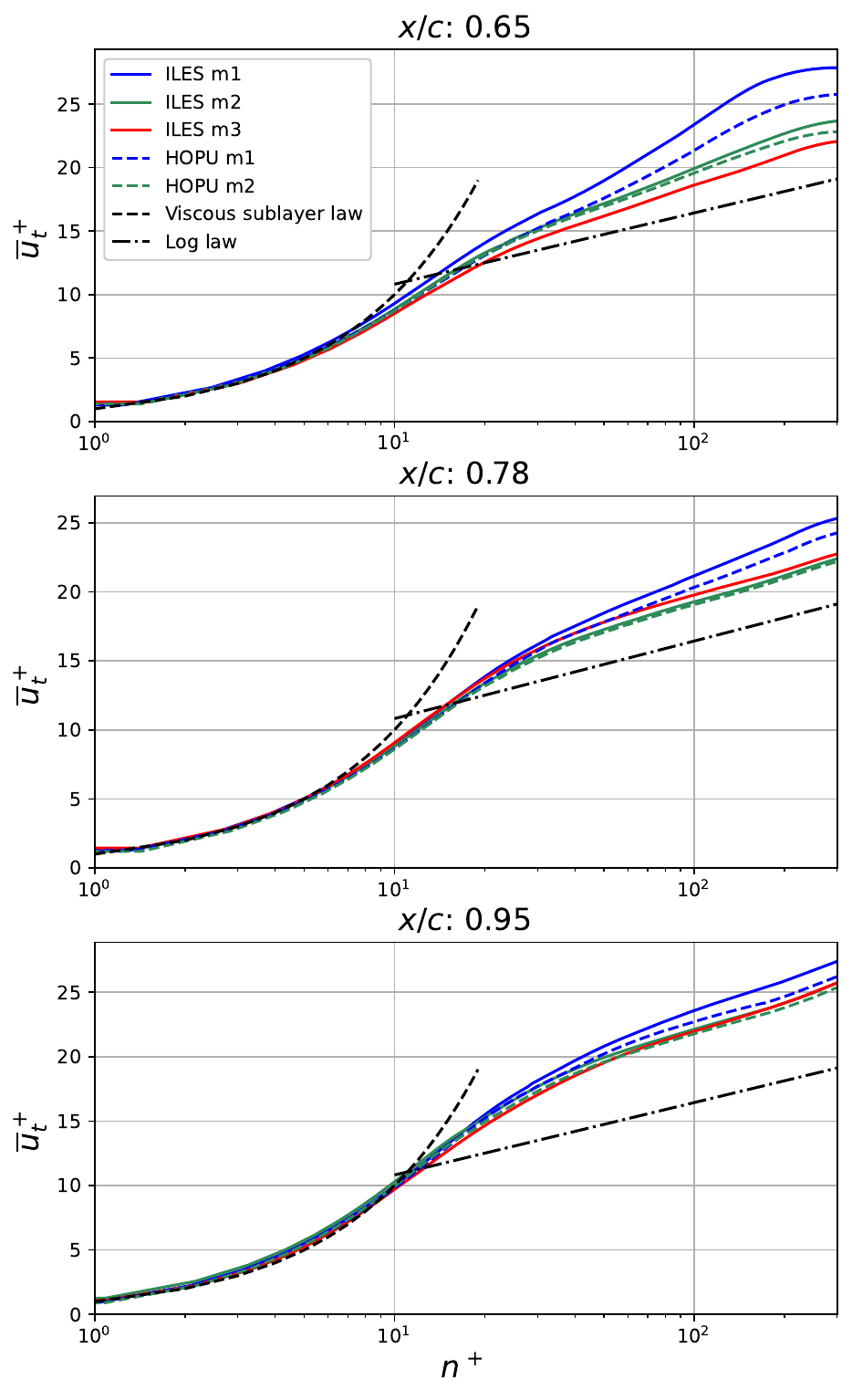}
	\caption{}
	\centering
	\label{fig:ut_a4}
\end{subfigure}    
\caption{Normalized mean tangential velocity $\overline{\underline{u}}_t^+$ at given locations along the upper surface for (a) $R1/\alpha 10$ and (b) $R3/\alpha 4$.}
\label{fig:ut}
\end{figure}

\begin{figure}
\centering
\begin{subfigure}{0.4\textwidth}
    \includegraphics[width=\textwidth]{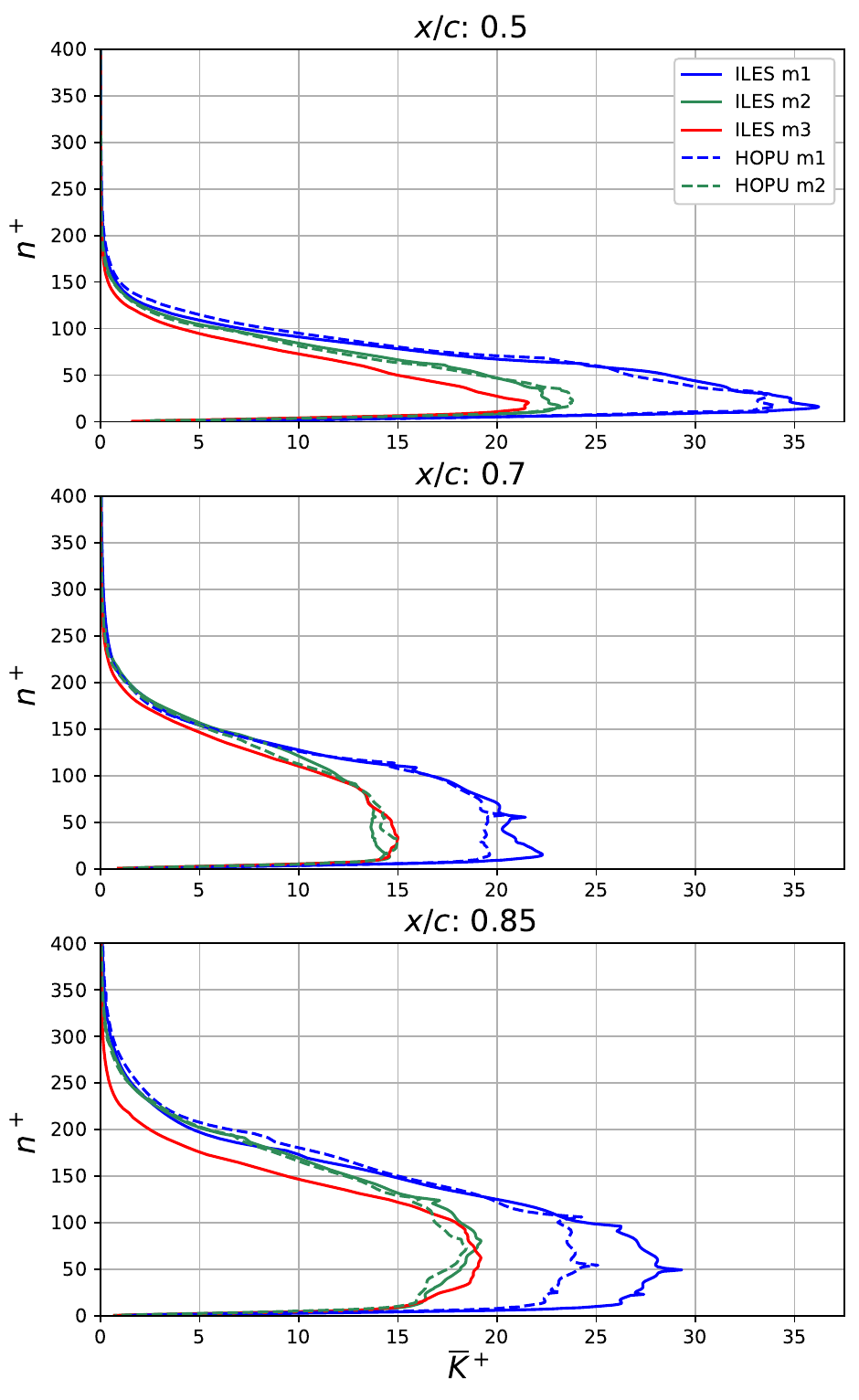}
    \caption{}
    \label{fig:k_a10}
\end{subfigure}
~
\begin{subfigure}{0.4\textwidth}
    \includegraphics[width=\textwidth]{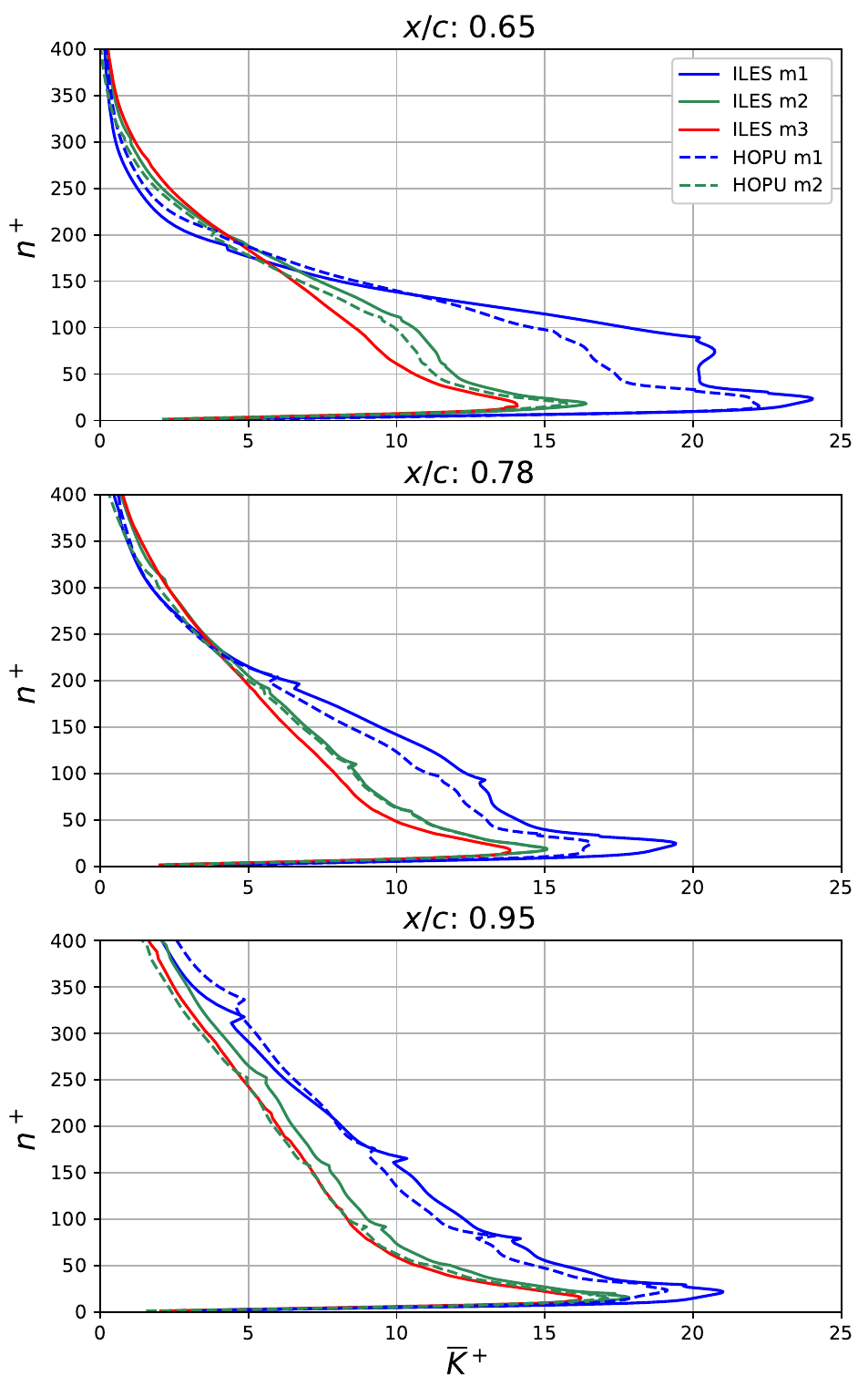}
	\caption{}
	\centering
	\label{fig:k_a4}
\end{subfigure}    
\caption{Normalized mean turbulent kinetic energy $\overline{K}^+$ at given locations along the upper surface for (a) $R1/\alpha 10$ and (b) $R3/\alpha 4$.}
\label{fig:k}
\end{figure}

\begin{figure}
\centering
\begin{subfigure}{0.4\textwidth}
    \includegraphics[width=\textwidth]{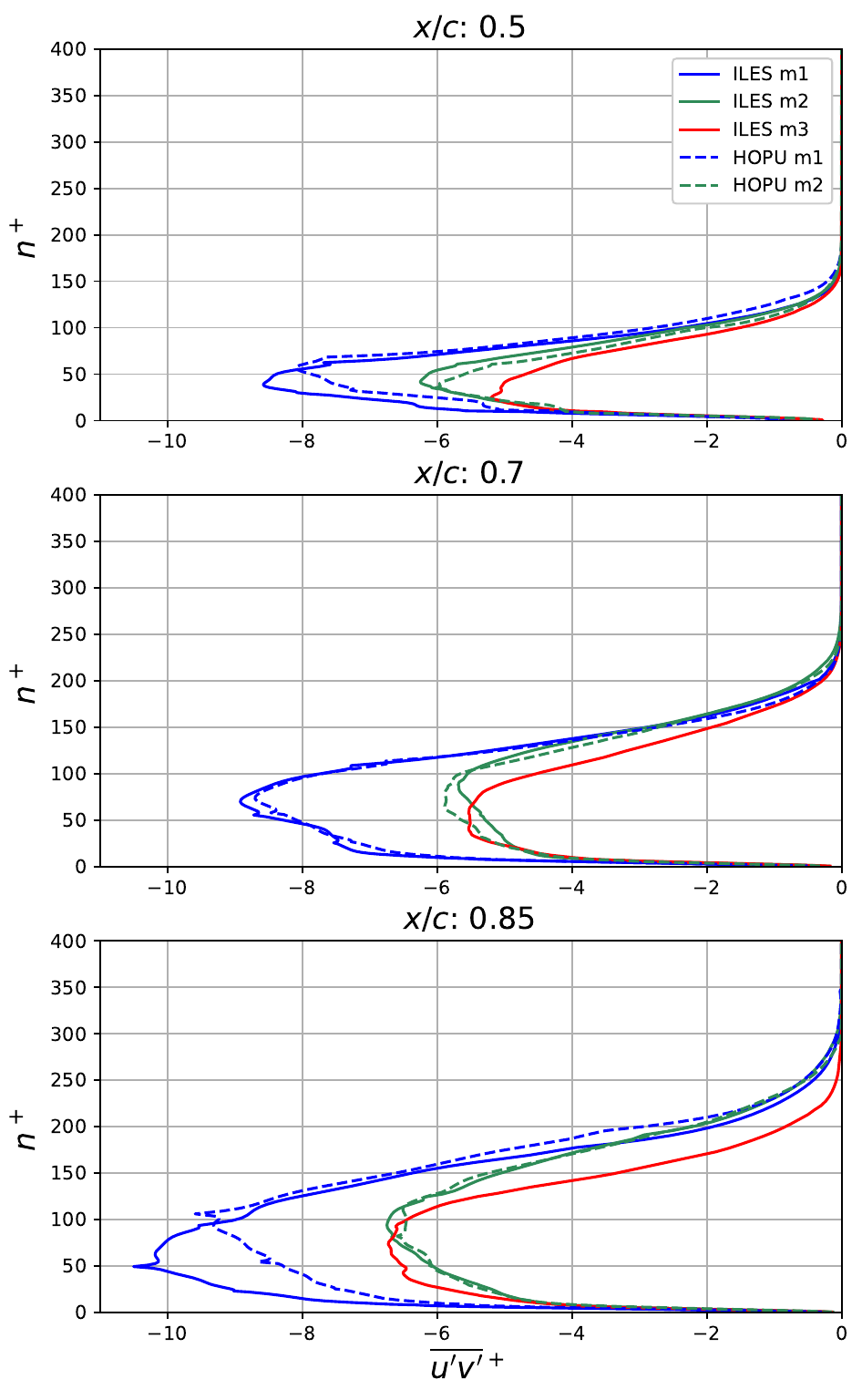}
    \caption{}
    \label{fig:uv_a10}
\end{subfigure}
~
\begin{subfigure}{0.4\textwidth}
    \includegraphics[width=\textwidth]{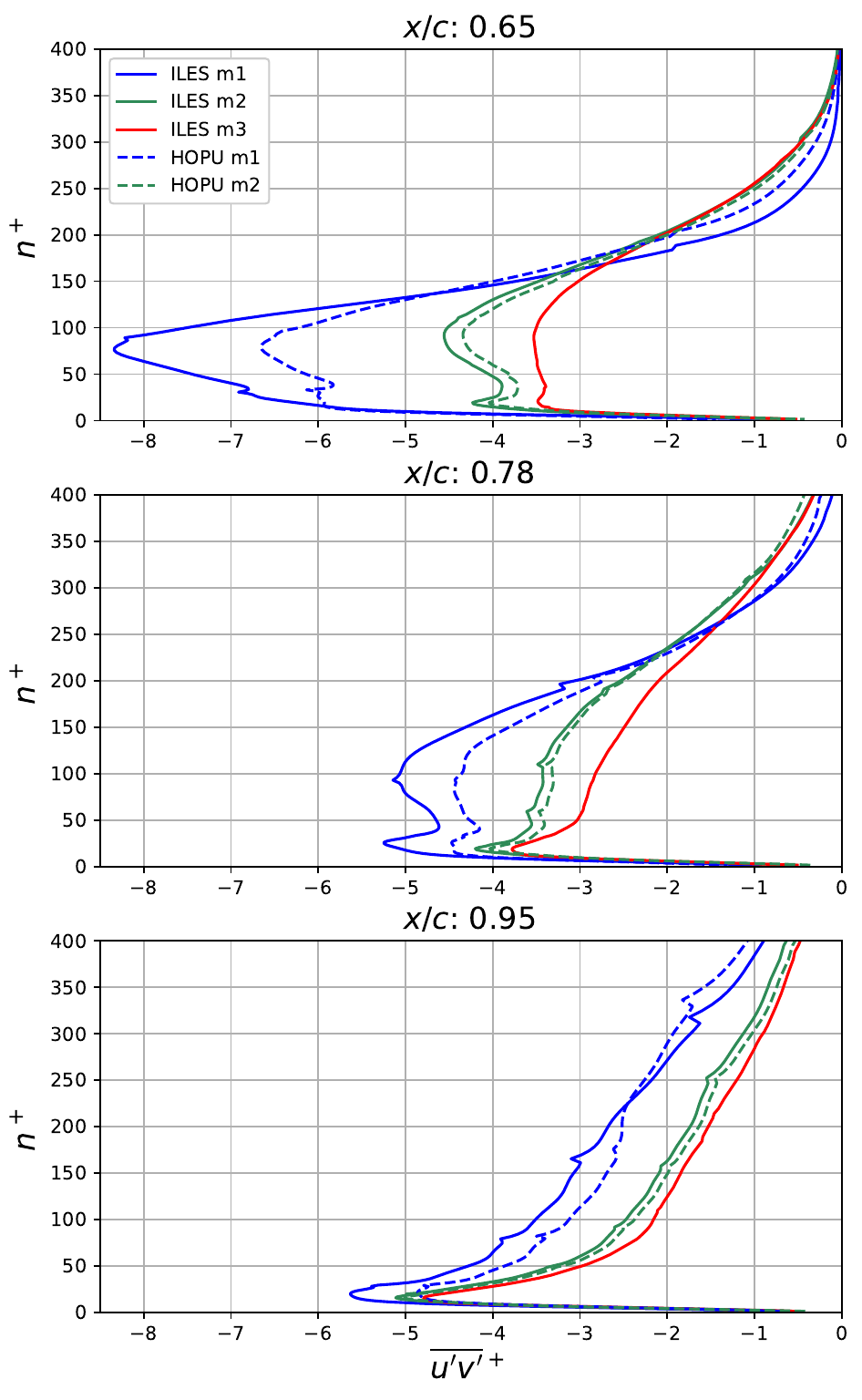}
	\caption{}
	\centering
	\label{fig:uv_a4}
\end{subfigure}    
\caption{Normalized mean Reynolds stress $\overline{u'v'}^+$ at given locations along the upper surface for (a) $R1/\alpha 10$ and (b) $R3/\alpha 4$.}
\label{fig:uv}
\end{figure}

The non-dimensional velocity profiles for the inner layer at different
locations along the turbulent portion of the flow are displayed in
Figure \ref{fig:ut}. The viscous sublayer given by
$|\overline{\underline{u}}_t|^+=n^+$ and the logarithmic layer law
$|\overline{\underline{u}}_t|^+=(1/\kappa) log (n^+) + C^+$ with
using $\kappa = 0.41$ and $C^+=5.2$ for the constant parameters are also
shown. In general, the logarithmic layer ($40 < n^+ < 200$) strongly
depends on the pressure gradient and the used parameters are experimentally fitted for mild
adverse pressure gradients. Note that the local friction
Reynolds number, depending on the friction velocity $u_{\tau}$, is 
calculated in order to obtain the dimensionless quantities.
The measurement locations in Figure \ref{fig:ut},
\ref{fig:k} and \ref{fig:uv} for $R1/\alpha10$ and $R3/\alpha4$ differ because of the varying points of reattachment and the subsequent start of the turbulent boundary layer.  
Overall, it is observed across all test cases in Figure \ref{fig:ut},
\ref{fig:k} and \ref{fig:uv} that increasing spatial resolution leads to convergence of results towards
those obtained with the finest mesh. Remarkably, even the highly
under-resolved settings yield meaningful predictions of the involved quantities.
In the subsequent discussion, we analyse the results in more detail.
The viscous sublayer ($n^+ < 5$) for $|\overline{\underline{u}}_t|^+$ is accurately approximated by all test cases.
However, significant distinctions between ILES and HOPU (particularly with mesh m1) are observed in the region $n^+ > 10$.
Here, compared to the ILES, the HOPU method tends to overpredict less and provides a better approximation of the velocity
profiles as shown in Figure \ref{fig:ut}. 
Due to the less restrictive upwind term in HOPU, which selectively penalizes only a subset of the
tangential velocity to maintain stability, the corresponding velocity profiles exhibit an
under-dissipative behaviour. Similar observations were previously reported in \cite{HOPU2023} for wall-bounded turbulent flows. 
When comparing $R1/\alpha10$ to $R3/\alpha4$, the velocity profiles of HOPU show that the $R1/\alpha10$ case provides significantly better improvements relative to the ILES with the same problem size.
In particular, the HOPU m1 case performs almost as well as the ILES m3 in terms of velocity profiles, as shown in the left 
column in Figure \ref{fig:ut}.
HOPU seems to perform better for $R1/\alpha10$, which can be attributed to the larger extent of the turbulent flow in this case. The increased turbulent activity likely enhances the effectiveness of HOPU, leading to improved results compared to the other scenario.

In Figure \ref{fig:k} and \ref{fig:uv}, the dimensionless mean 
turbulent kinetic energy $\overline{K}^+$ and mean Reynolds stress
component $\overline{u'v'}^+$ are shown. Comparing HOPU
with ILES, similar to our previous observations, the results obtained with the projected
upwind variant are less pronounced to overprediction than with the standard
method. This can be seen across most locations for $\overline{K}^+$ and
$|\overline{\underline{u}}_t|^+$. 

\begin{figure}[t]
\centering
\includegraphics[width=0.7\textwidth]{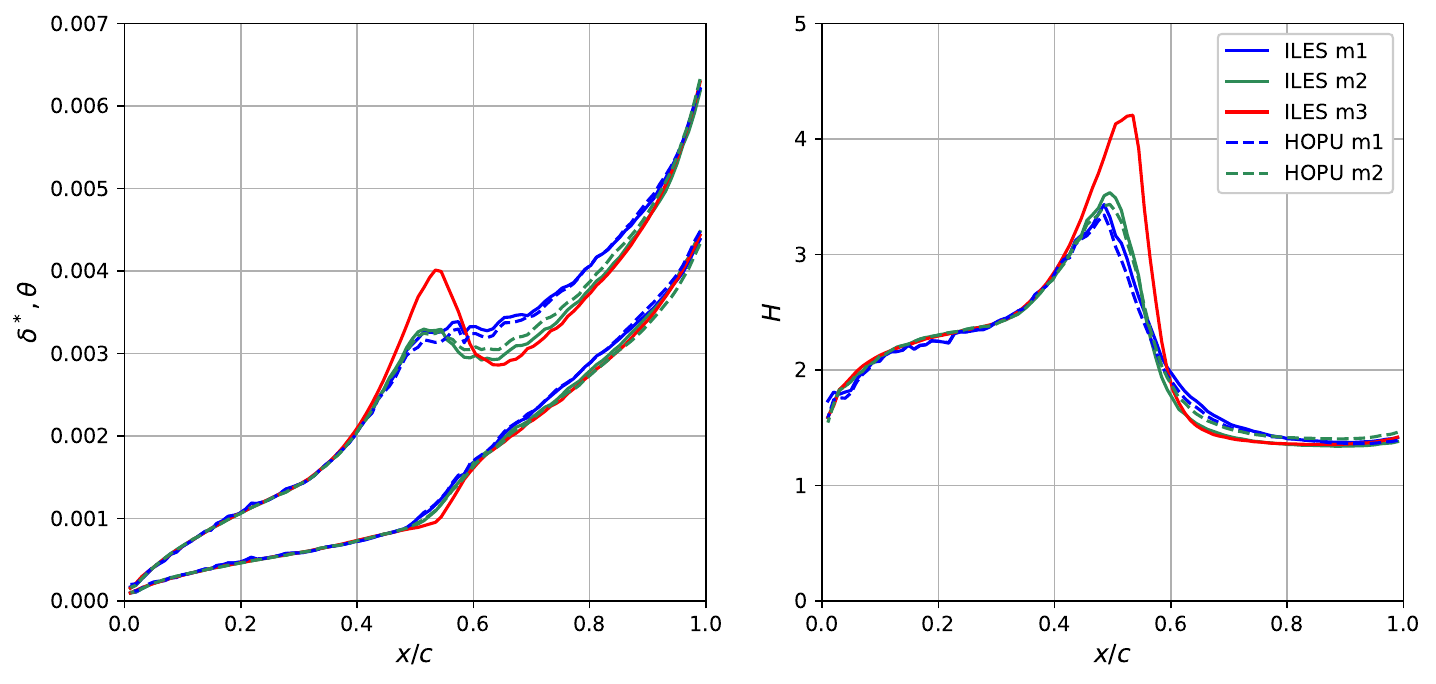}
\caption{Streamwise displacement $\delta^*$ (left, upper curve) and momentum thickness $\theta$ (left, lower curve) and shape parameter $H$ (right) along the upper surface for $R3/\alpha4$.}
\centering
\label{fig:bl}
\end{figure}

Boundary layer separation leads to a rapid increase in displacement
thickness $\delta^*$ and thus the shape parameter $H$, as indicated in Figure
\ref{fig:bl}. Upon reaching the end of the separation bubble, $\delta^*$ decreases
as the boundary layer reattaches. A strong pressure gradient induces an increase in both boundary layer thicknesses towards the trailing edge.
In the laminar portion of the boundary layer flow, all test cases show
good agreement. However, the sharp peak in $\delta^*$ during separation and its subsequent 
reduction is not well captured by the highly under-resolved ILES cases
m1 and m2. The HOPU variant shows negligible differences to the standard method. 

%% file: chapters/summary.tex
\section{Discussion and Conclusion}\label{conclusion}

A scalable formulation of the MCS approach for solving structure-resolving
turbulent flow has been evaluated for the Eppler 387 airfoil test case.
Utilizing a projection scheme in conjunction with
preconditioned conjugate gradient solver, we have described a computational
efficient implementation of the discrete MCS method.

The numerical results obtained with the finest grid in this study demonstrate
precise prediction of transition phenomena for Reynolds numbers of
$1 \cdot 10^5$ ($\alpha = 10 \degree$) and $3 \cdot 10^5$ ($\alpha = 4 \degree$)
compared to 
experimental data. This underscores the advantage of the low-dissipative
mechanism of MCS in simulating transition, as it effectively preserves the
small-scale instabilities without excessive damping.

In addition to the standard implicit approach, the results obtained 
with the high-order projected upwind method provide a more refined insight into how
numerical dissipation effects the transition from laminar to turbulent flow.
While HOPU demonstrates similar behaviour in approximating the
transition process, it improves the boundary layer prediction in the 
turbulent portion of the flow. We infer here that rising disturbances in
the laminar boundary layer, which lead to transition, are not 
significantly affected by the dissipation from the more refined
convection stabilization. Additionally, according to the given
outcomes it can be inferred that HOPU performs
well in wall-bounded turbulent flows regardless of the transition
mechanism. Comparing the two different flow scenarios, we conclude
that HOPU gives more profound improvements to ILES for $R1/\alpha10$ for the velocities.
This may reason in the larger extent of turbulent wall-bounded flow compared to $R3/\alpha4$.

It is currently uncertain if utilizing the absolute value jump $\eta$ as an indication is the best strategy, and further study on the adaptive
algorithm is required. Future research will focus on the improvement of
this upwind variant and extension to \xaver{scenarios with fully turbulent
flows at high Reynolds numbers}.